%% file: main.tex
\documentclass[aps,prl,reprint,groupedaddress]{revtex4-2}

\usepackage{graphicx} 
\usepackage{amsmath} 
\usepackage{amssymb} 
\usepackage{braket}
\usepackage{physics} 
\usepackage{xcolor} 
\usepackage{mathtools} 
\usepackage{slashed} 
\usepackage{parskip}
\usepackage{comment}
\usepackage{standalone}

\renewcommand{\eqref}[1]{Eq.~(\ref{#1})}

\newcommand{\figref}[1]{Fig.~\ref{#1}}
\renewcommand{\b}[1]{\textbf{#1}}
\renewcommand{\rm}[1]{\textrm{#1}}

\usepackage{chngcntr}
\counterwithout{equation}{section}

\newcommand{\pati}[1]{}

\begin{document}


\title{Photon State Evolution in Arbitrary Time-Varying Media}


\author{Artuur Stevens}
\email[]{artuur.stevens@kuleuven.be}

\author{Christophe Caloz}
\affiliation{KU Leuven}


\date{\today}

\begin{abstract}
We introduce the instantaneous eigenstate method to study the evolution of quantum states in non-dispersive, homogeneous media with arbitrary time-varying permittivity and permeability. This method leverages the Heisenberg equation to bypass the Schrödinger equation, which leads to a complicated infinite set of coupled differential equations. Instead, the method allows the computation of the state evolution by solving only two coupled differential equations. Using this approach, we draw general conclusions about photon statistics in time-varying media. Our findings reveal that the maximum probability of generating a single photon pair from vacuum in such media is $25\%$, while Bell states can be created with a maximum probability of $84\%$. 
Additionally, we demonstrate that the spectral response of emitted photons depends on the temporal profiles of permittivity and permeability and provide a method to determine the required profiles for specified spectral profiles.
These results provide deep insights into photon state manipulation in time-varying media. Furthermore, the instantaneous eigenstate method opens new opportunities to study state evolution in other systems where the Heisenberg equation offers a more tractable solution than the Schrödinger equation.
\end{abstract}


\maketitle

Time-varying media represent a new frontier in quantum optics, where dynamically manipulating material properties opens doors to unprecedented physical phenomena and applications.
On the one hand, the energy non-conservation inherent in time-varying media gives rise to interesting phenomena, such as the creation and annihilation of photon pairs either through modulating the permittivity and permeability~\cite{louisell_quantum_1961, gordon_quantum_1963, mendonca_quantum_2000, Horsley_Pendry_2023} or via time-dependent boundaries, as in the dynamic Casimir effect~\cite{moore_quantum_1970, HUI2000174, dodonov_dynamical_2009, roman-ancheyta_dynamical_2017, ma_enhanced_2019, li_preparation_2023}.
On the other hand, advances in technology have made the experimental realization of time-varying media possible, 
as demonstrated in microwave transmission lines~\cite{wilson_observation_2011, moussa_observation_2023, jones_time-reflection_2024}
and highly-doped semiconductors~\cite{caspani_enhanced_2016, zhou_broadband_2020, bohn_spatiotemporal_2021, tirole_double-slit_2023, galiffi_optical_2024}.
These developments have driven innovations in quantum technologies, enabling applications such as anti-reflection temporal coatings~\cite{pacheco-pena_antireflection_2020, liberal_quantum_2023}, which minimize quantum noise, 
enhanced lasing techniques~\cite{dikopoltsev_light_2022, lyubarov_amplified_2022, lyubarov_controlling_2024}, directional amplification for optimized signal propagation~\cite{vazquez-lozano_shaping_2023} and quantum squeezing~\cite{ganfornina-andrades_quantum_2023}.

One of the most intriguing aspects of time-varying media is the creation of photon pairs from vacuum. 
This property makes them particularly suitable as heralded single-photon sources, where the detection of one photon in the pair signals the presence of the other, enabling applications in quantum communication and cryptography.
In such applications, it is important to understand how many photon pairs will be formed and how the properties of the time-varying medium influence this process.
Photon pair production in linear media with a sudden, spatially uniform change in permittivity, known as a temporal interface, has already been studied~\cite{mendonca_quantum_2000, mendonca_temporal_2003, mendonca_time_2005, mirmoosa_quantum_2025}. 

So far, photon pair production statistics have not been studied for media with general temporal profiles of the permittivity and permeability.
To calculate photon pair production probabilities, one typically solves the Schrödinger equation, which governs the time evolution of the quantum state.
However, in time-varying system, the Schrödinger equation becomes an infinite set of coupled differential equations with time-dependent coefficients, 
making it analytically and computationally challenging to solve.
Moreover, even if the equations are solved for a specific temporal profile, they fail to provide broader insights into important properties of general time-varying media,
such as the maximum probability of generating a single photon pair, if there is such a limit.
Consequently, the Schrödinger equation offers only a limited understanding of the possibilities for photon pair production in time-varying media.

In this paper, we introduce a new method, the instantaneous eigenstate method, to calculate the evolution of quantum states in non-dispersive, homogeneous, arbitrarily time-varying media.
This method leverages the Heisenberg equations to determine a state's evolution as a function of two complex functions, which are solutions to two coupled differential equations, and depend on the modulation profiles.
We can use the related functions to draw general conclusions about the possibilities of photon state manipulation using time-varying media.

The evolution of a quantum state, $\ket{\psi(t)}$, is governed by the Schrödinger equation, $i\hbar \partial_t \ket{\psi(t)}  = \hat{H}(t) \ket{\psi(t)}$, where $\hat{H}(t)$ is the Hamiltonian of the system.
In a stationary, linear and isotropic medium with permittivity $\epsilon_1$ and permeability $\mu_1$, confined within a conducting cavity of volume $V$, this Hamiltonian may be expressed as
\begin{equation}\label{Eq.: Stationary Hamiltonian}
    \begin{aligned}
         \hat{H}(t) &= \frac{1}{2} \int_V  \rm{d}^3 r \left(  \frac{\hat{\b{D}}^2(\b{r}, t)}{\epsilon_1}   +  \frac{(\curl{\hat{\b{A}}}(\b{r}, t))^2}{\mu_1}   \right),
    \end{aligned}
\end{equation}
where $\hat{\b{A}}$ is the vector potential and $\hat{\b{D}}$ is the electric displacement field~\cite{hillery_introduction_2009}.
These fields are operators satisfying $[\hat{A}_i(\b{r}, t), -\hat{D}_j(\b{r}', t)] = i\hbar \delta^\rm{T}_{ij}(\b{r}-\b{r}')$, where $\delta^\rm{T}_{ij}$ is the transverse delta function \cite{cohen-tannoudji_photons_2024, loudon_quantum_2000}. The field operators $\hat{\b{A}}$ and $\hat{\b{D}}$ can be expanded in Fourier series as
\begin{equation}\label{Eq.: A and D expansion}
    \begin{aligned}
     \hat{\b{A}}(\b{r}, t)  &=\sum_{\b{k}, \lambda} \sqrt{\frac{\hbar}{2\epsilon_1 \omega_k  V}} \b{e}_{\b{k},\lambda} \left( \hat{a}_{\b{k}, \lambda}(t)e^{i\b{k}\cdot\b{r}} +  \hat{a}^\dagger_{\b{k}, \lambda}(t)e^{-i\b{k}\cdot\b{r}}\right), \\
     \hat{\b{D}}(\b{r}, t)  &=\sum_{\b{k}, \lambda}i \sqrt{\frac{\hbar \epsilon_1 \omega_k}{2V}} \b{e}_{\b{k},\lambda} \left(   \hat{a}_{\b{k}, \lambda}(t) e^{i\b{k}\cdot\b{r}}-  \hat{a}^\dagger_{\b{k}, \lambda}(t)e^{-i\b{k}\cdot\b{r}}\right),\\    
    \end{aligned}
\end{equation}
which define the creation and annihilation operators, $\hat{a}^\dagger_{\b{k}, \lambda}$ and $\hat{a}_{\b{k}, \lambda}$, respectively, and where $\lambda$ is the polarization index, $\b{e}_{\b{k},\lambda}$ are the polarization vectors, and $\omega_k = |\b{k}|/\sqrt{\epsilon_1 \mu_1}$ is the mode frequency.
In addition, in the cavity, assumed of length $L$, the wave vectors are quantized as $\b{k} = \frac{2\pi}{L} (n_x, n_y, n_z)$, where $n_x$, $n_y$ and $n_z$ are integers.
Finally, substituting the Fourier expansions of \eqref{Eq.: A and D expansion} into \eqref{Eq.: Stationary Hamiltonian} yields the usual form of the Hamiltonian, $\hat{H}(t) = \sum_{\b{k}, \lambda} \hbar \omega_k \left(\hat{n}_{\b{k}, \lambda}(t) +1/2     \right)$,
where $\hat{n}_{\b{k}, \lambda}(t) = \hat{a}^\dagger_{\b{k}, \lambda}(t) \hat{a}_{\b{k}, \lambda}(t)$ is the photon number operator. The eigenstates of the number operator, $\hat{n}_{\b{k}, \lambda}(t)$, define the Fock states, $\ket{n}_{\b{k}, \lambda}$, with $n$ photons in the ($\b{k}, \lambda$) wave mode. These states are then also eigenstates of the Hamiltonian of the medium, which implies that the number of photons is conserved.

As the permittivity and permeability of the medium begin to vary in time, the Hamiltonian of the system changes accordingly.
As shown in Sec.~$1$ of \cite{Supplementary_material}, the Hamiltonian retains the same form as in \eqref{Eq.: Stationary Hamiltonian}, 
with the constant permittivity and permeability replaced by their time-dependent counterparts, so that \cite{stevens_photon_2024}
\begin{equation}
    \begin{aligned}
         \hat{H}(t) &= \frac{1}{2} \int_V  \rm{d}^3 r \left(  \frac{\hat{\b{D}}^2(\b{r}, t)}{\epsilon(t)}   +  \frac{(\curl{\hat{\b{A}}}(\b{r}, t))^2}{\mu(t)}   \right).
    \end{aligned}
\end{equation}
Substituting \eqref{Eq.: A and D expansion} into this Hamiltonian, we obtain
\begin{equation}\label{Eq.: H(a, ad, t)}
    \begin{aligned}
     \hspace{-2.8mm} \hat{H}(t) &= \sum_{\bf{k}, \lambda} \hbar \alpha_k(t) \left(   \hat{n}_{\bf{k}, \lambda}(t) + \frac{1}{2}     \right) \\
     &\quad + \frac{\hbar \beta_k(t)}{2} \left(\hat{a}_{\bf{k}, \lambda}(t)  \hat{a}_{-\bf{k}, \lambda}(t)
      +
     \hat{a}^\dagger_{\bf{k}, \lambda}(t) \hat{a}^\dagger_{-\bf{k}, \lambda}(t) \right) \hspace{-0.75mm}.\\
    \end{aligned}
\end{equation}
The first term represents the energy per photon and defines the instantaneous frequency \mbox{$\alpha_k(t) = \frac{\omega_k}{2}\left(\frac{\epsilon_1}{\epsilon(t)} + \frac{\mu_1}{\mu(t)}\right)$}.
The second term describes the creation and annihilation of photon pairs of the same polarization but opposite momenta,
with interaction strength $\beta_k(t) = \frac{(-1)^\lambda \omega_k}{2}\left(\frac{\epsilon_1}{\epsilon(t)} - \frac{\mu_1}{\mu(t)}\right)$.
These interactions correspond to pairwise creation and annihilation events and ensure the conservation of momentum, while the energy can change in time~\cite{morgenthaler_velocity_1958, caloz_spacetime_2019, koutserimpas_electromagnetic_2020, ortega-gomez_tutorial_2023, galiffi_photonics_2022}.
The Hamiltonian in \eqref{Eq.: H(a, ad, t)} has the same form as that used in non-degenerate parametric down-conversion \cite{BoydRobertW.2008NO(E, scully_quantum_1997}, but with time-dependent coefficients $\alpha_k(t)$ and $\beta_k(t)$ without requiring the non-depletion approximation.

To study the evolution of an initial state $\ket{\psi(0)}$, we expand the state at later times as $\ket{\psi(t)} = \sum_{n,m=0}^\infty C_{n, m}(t) \ket{n, m}$.
While the coefficients $C_{n, m}(t)$ would normally be calculated using the Schrödinger equation, we show in Sec.~$2$ of~\cite{Supplementary_material} that this would lead to an infinite set of coupled differential equations with time-varying coefficients, which is both analytically difficult and computationally intensive. 
Instead, we will make use of the unitary time-evolution operator $\hat{U}(t)$, so that $\ket{\psi(t)} = \hat{U}(t)\ket{\psi(0)}$.
For the orthonormal Fock states, we have $\bra{n',m'} \ket{n,m} = \delta_{n, n'} \delta_{m, m'}$, so the coefficients $C_{n,m}(t)$ are found as $C_{n, m}(t) = \bra{n,m}\hat{U}(t)\ket{\psi(0)}$.
The time-evolution operator is given by $\hat{U}(t) = T\left\{\exp\left((i\hbar)^{-1}\int_0^t \hat{H}(\tau) d\tau \right) \right\}$, where $T$ denotes the time-ordering operator.
However, a closed-form expression for this operator is difficult to obtain, as it involves the Dyson series~\cite{dyson_s_1949, sakurai_modern_2020}.
Instead, we define the state $\ket{\xi_{n, m}(t)} = \hat{U}^\dagger(t)\ket{n,m}$, which allows us to express the coefficients as $C_{n, m}(t) = \bra{\xi_{n, m}(t)}\ket{\psi(0)}$.
This approach allows to calculate the coefficients by determining the states $\ket{\xi_{n, m}(t)}$, which, as we will show, does not require the explicit determination of the time-evolution operator.

The states $\ket{\xi_{n, m}(t)}$ possess interesting properties that simplify their derivation.
In the Heisenberg picture, an operator $\hat{O}(t)$ would evolve as $\hat{O}(t) = \hat{U}^\dagger(t) \hat{O}(0) \hat{U}(t)$.
As detailed in Sec.~$3$ of \cite{Supplementary_material}, the unitarity of $\hat{U}(t)$ implies that $\ket{\xi_{n, m}(t)} = \hat{U}^\dagger(t)\ket{n,m}$ satisfies
\begin{equation}\label{Eq.: Instantaneous Relations}
   \left\{
       \begin{aligned}
        \hat{n}_{\b{k}}(t) \ket{\xi_{n, m}(t)} &= n \ket{\xi_{n, m}(t)},\\
        \hat{a}_{\b{k}}(t) \ket{\xi_{n, m}(t)} &= \sqrt{n} \ket{\xi_{n-1, m}(t)},\\
        \hat{a}^\dagger_{\b{k}}(t) \ket{\xi_{n, m}(t)} &= \sqrt{n+1} \ket{\xi_{n+1, m}(t)},
       \end{aligned}
   \right.
\end{equation}
and similar relations for the backward operators $\hat{n}_{-\b{k}}(t)$, $\hat{a}_{-\b{k}}(t)$ and $\hat{a}^\dagger_{-\b{k}}(t)$.
The first equation indicates that the state $\ket{\xi_{n, m}(t)}$ is an eigenstate of the time-dependent number operator $\hat{n}_{\b{k}}(t)$, or an \textit{instantaneous eigenstate}.
The second equation implies the existence of an instantaneous ground state $\ket{\xi_{0, 0}(t)}$, which satisfies $\hat{a}_{\pm \b{k}}(t) \ket{\xi_{0, 0}(t)} = 0$.
Finally, the third equation allows the construction of all instantaneous eigenstates by repeatedly applying the time-dependent creation operators to the instantaneous ground state,
yielding
\begin{equation}
   \begin{aligned}
    \ket{\xi_{n, m}(t)} = \frac{\left(\hat{a}^\dagger_{\b{k}}(t) \right)^n \left(\hat{a}^\dagger_{-\b{k}}(t) \right)^m}{\sqrt{n! m!}}\ket{\xi_{0, 0}(t)}.
   \end{aligned}
\end{equation}
Taking the Hermitian conjugate of this equation and using $C_{n, m}(t) = \bra{\xi_{n, m}(t)}\ket{\psi(0)}$, we obtain
\begin{equation}\label{Eq.: Coefficient formula}
    \begin{aligned}
        C_{n, m}(t) = \frac{1}{\sqrt{n! m!}} \left\langle \xi_{0, 0}(t) \left| \hat{a}^m_{-\b{k}}(t) \hat{a}^n_{\b{k}}(t) \right| \psi(0) \right\rangle.
    \end{aligned}
\end{equation}
Next, to fully determine the coefficients $C_{n, m}(t)$, we will calculate the time-dependent annihilation operators $\hat{a}_{\pm \b{k}}(t)$ and the instantaneous ground state $\ket{\xi_{0, 0}(t)}$.

The time-dependent annihilation operators $\hat{a}_{\pm \b{k}}(t)$ follow the Heisenberg equations $i\hbar \partial_t \hat{a}_{\pm \b{k}}(t) = [\hat{a}_{\pm \b{k}}(t), \hat{H}(t)]$.
The commutation relations $[\hat{a}_{\b{k}}(t), \hat{a}^\dagger_{\b{k}'}(t)] = \delta_{\b{k}, \b{k}'}$ and $[\hat{a}_{\b{k}}(t), \hat{a}_{\b{k}'}(t)] = [\hat{a}^\dagger_{\b{k}}(t), \hat{a}^\dagger_{\b{k}'}(t)] = 0$ imply that~\cite{stevens_photon_2024}
\begin{equation}\label{Eq.: Differential equations a}
   \begin{aligned}
        i\hbar \partial_t \hat{a}_{\pm \b{k}}(t) = \alpha_k(t) \hat{a}_{\pm \bf{k}}(t) + \beta_k(t) \hat{a}^\dagger_{\mp \bf{k}}(t).
   \end{aligned}
\end{equation}
Since the medium is isotropic, we expect the solution for $\hat{a}_{+ \b{k}}(t)$ and $\hat{a}_{- \b{k}}(t)$ to be similar.
Thus, we take the ansatz
\begin{equation}\label{Eq.: Ansatz}
   \begin{aligned}
        \hat{a}_{\pm \b{k}}(t) = f_k(t)\hat{a}_{\pm \b{k}} + g_k(t)\hat{a}^\dagger_{\mp \b{k}}, 
   \end{aligned}
\end{equation}
where $\hat{a}_{\pm \b{k}}$ and $\hat{a}^\dagger_{\mp \b{k}}$ are the operators defined at $t=0$, right before the modulation starts, so $f_k(0) = 1$ and $g_k(0) = 0$.
This ansatz is the well-known Bogoliubov transformation \cite{bennemann_superconductivity_2008} with the time-dependent coefficients $f_k(t)$ and $g_k(t)$ that describe the evolution of the annihilation operators in time-varying media.
Since the Fock states, $\ket{n, m}$, are also defined before the modulation, the initial operators act as $\hat{a}_{+ \b{k}}\ket{n, m} = \sqrt{n}\ket{n-1, m}$ and $\hat{a}^\dagger_{+ \b{k}}(t)\ket{n, m} = \sqrt{n+1}\ket{n+1, m}$.
We substitute the ansatz of~\eqref{Eq.: Ansatz} into the Heisenberg equations~\eqref{Eq.: Differential equations a} and compare terms proportional to the same initial operators, 
and find that the functions $f_k$ and $g_k$ satisfy
\begin{equation}\label{Eq.: Differential equations f and g}
    \left\{
        \begin{aligned}
         \partial_t f_k(t) &= -i\alpha_k(t)f_k(t) -i  \beta_k(t)g_k^*(t),\\
         \partial_t g_k(t) &= -i\alpha_k(t)g_k(t) -i \beta_k(t)f^*_k(t).\\
        \end{aligned}
    \right.
\end{equation}
These differential equations can then be solved analytically for some specific temporal modulations, such as temporal steps, as shown in Sec.~$4$ of \cite{Supplementary_material}, or impedance-matched modulations, where $\beta_k(t)$ vanishes and $f_k(t)$ becomes a phase factor with $g_k(t) = 0$.
For more general modulation profiles, they can be solved numerically.

The only unknown left in \eqref{Eq.: Coefficient formula} is the instantaneous ground state, $\ket{\xi_{0, 0}(t)}$, which satisfies $\hat{a}_{\pm \b{k}}(t) \ket{\xi_{0, 0}(t)} = 0$.
First, we expand this state in the Fock state basis as $\ket{\xi_{0, 0}(t)} = \sum_{n,m} D_{n, m}(t) \ket{n, m}$.
Using the conditions $\hat{a}_{\pm \b{k}}(t) \ket{\xi_{0, 0}(t)} = 0$ and \eqref{Eq.: Ansatz}, we show in Sec.~$4$ of \cite{Supplementary_material} that $D_{n, m}(t)$ satisfies
\begin{equation}
   \begin{aligned}
        D_{n, m}(t) = \left(-\frac{g_k(t)}{f_k(t)}\right)^n D_{0,0}(t) \delta_{n, m}.
   \end{aligned}
\end{equation}
Since the instantaneous eigenstate must be normalized, $\bra{\xi_{0, 0}(t)}\ket{\xi_{0, 0}(t)} = 1$, we find that $D_{0, 0}(t)$ is determined up to a phase factor $\theta(t)$, so that $D_{0, 0}(t) = e^{i\theta(t)}/f_k(t)$.
Therefore, we conclude that the instantaneous ground state $\ket{\xi_{0, 0}(t)}$ is given by 
\begin{equation}\label{Eq.: Instantaneous Ground state}
   \begin{aligned}
    \ket{\xi_{0, 0}(t)} = \frac{e^{i\theta(t)}}{f_k(t)} \sum_{n} \left(-\frac{g_k(t)}{f_k(t)}\right)^n \ket{n, n}.
   \end{aligned}
\end{equation}

We can now substitute \eqref{Eq.: Ansatz} and \eqref{Eq.: Instantaneous Ground state} into \eqref{Eq.: Coefficient formula} to calculate the coefficients $C_{n, m}(t)$ up to a phase factor $\theta(t)$. 
As an example, we calculate in Sec.~$5$ of \cite{Supplementary_material} that for the initial state $\ket{0, 0}$, the coefficients $C_{n, m}(t)$ are given by
\begin{equation}\label{Eq.: Vacuum Coefficients}
    \begin{aligned}
        C_{n, m}(t) = \frac{g^n_k(t)}{(f_k^{n+1}(t))^*}e^{-i\theta(t)} \delta_{n, m}.
    \end{aligned}
\end{equation}
To determine the phase factor $\theta(t)$, we can insert \eqref{Eq.: Vacuum Coefficients} into the Schrödinger equation.
As detailed in Sec.~$5$ of \cite{Supplementary_material}, we find that the coefficients satisfy the Schrödinger equation only if $\theta(t) = 0$.
Therefore, the initial state $\ket{0,0}$ evolves into the state 
\begin{equation}\label{Eq.: Vacuum Evolution}
   \begin{aligned}
        \ket{\psi(t)} &= \sum_n \frac{g^n_k(t)}{(f_k^{n+1}(t))^*} \ket{n, n}, \\
   \end{aligned}
\end{equation}
with $f_k(t)$ and $g_k(t)$ determined by the differential equations \eqref{Eq.: Differential equations f and g}.
Since $\theta(t) = 0$, the instantaneous ground state \eqref{Eq.: Instantaneous Ground state} is now completely determined.
Together with \eqref{Eq.: Coefficient formula} and \eqref{Eq.: Ansatz}, the coefficients $C_{n, m}(t)$ can be calculated for any initial state in terms of $f_k(t)$ and $g_k(t)$.
Thus, instead of solving the Schrödinger equation for the coefficients $C_{n, m}(t)$, an infinite set of coupled differential equations, 
we only need to solve the two coupled differential equations in \eqref{Eq.: Differential equations f and g} for $f_k$ and $g_k$ to determine the coefficients $C_{n, m}(t)$.

We can now study the effect of the temporal modulation profile on the photon field.
If the initial state of the system is vacuum, $\ket{0, 0}$, then, at a later time $t$, the state has evolved into \eqref{Eq.: Vacuum Evolution}.
At this time, the probability of detecting $n$ photon pairs is given by $P_{n, n}(t) = |C_{n, n}(t)|^2 = \frac{|g_k(t)|^{2n}}{|f_k(t)|^{2n+2}}$, 
indicating that the probability depends on two degrees of freedom, $|f_k(t)|$ and $|g_k(t)|$.
However, for the Bogoliubov transformation of~\eqref{Eq.: Ansatz}, $|f_k(t)|^2 - |g_k(t)|^2 = 1$ due to the commutation relation $[\hat{a}_{\b{k}}(t), \hat{a}^\dagger_{\b{k}}(t)] = 1$ \cite{bogoljubov1958new}.
Consequently, the probability $P_{n, n}(t)$ is determined by a single degree of freedom, $|g_k(t)|$, as 
\begin{equation}\label{Eq.: P(t_0)}
   \begin{aligned}
        P_{n, n}(t) = \frac{|g_k(t)|^{2n}}{(1 + |g_k(t)|^2)^{n+1}}.
   \end{aligned}
\end{equation}
From this expression, we deduce that the maximum probability of producing a single photon pair---attained at $|g_k(t)| = 1$---is $25\%$. 
When the modulation leads to $|g_k(t)|>1$, the likelihood of generating multiple photon pairs increases, which in turn reduces the probability of producing only a single photon pair.
Interestingly, this is the same limit as in the specific case of a temporal step~\cite{mirmoosa_quantum_2025}, which our theory recovers as follows. A temporal step from $\epsilon_1$ and $\mu_1$ for $t\leq 0$ to $\epsilon_2$ and $\mu_2$ for $t>0$ results in 
\begin{equation}
   \begin{aligned}
        |g_k(t)| = \frac{1}{2}\left|\sqrt{\frac{\mu_r}{\epsilon_r}} - \sqrt{\frac{\epsilon_\rm{r}}{\mu_\rm{r}}}\right| \cdot \left|\sin(\frac{\omega_k t}{\sqrt{\epsilon_\rm{r} \mu_\rm{r}}})\right|,
   \end{aligned}
\end{equation} 
where $\epsilon_\rm{r} = \epsilon_2 / \epsilon_1$ and $\mu_\rm{r} = \mu_2 / \mu_1$, as shown in Sec.~$4$ of \cite{Supplementary_material}.
Therefore, according to~\eqref{Eq.: P(t_0)}, a temporal step can achieve the maximum pair-generation probability of $25\%$ provided that $\frac{1}{2}\left|\sqrt{\frac{\mu_\rm{r}}{\epsilon_\rm{r}}} - \sqrt{\frac{\epsilon_\rm{r}}{\mu_\rm{r}}}\right|>1$.

\begin{figure}[!ht]
    \centering
    \includegraphics[width=\linewidth]{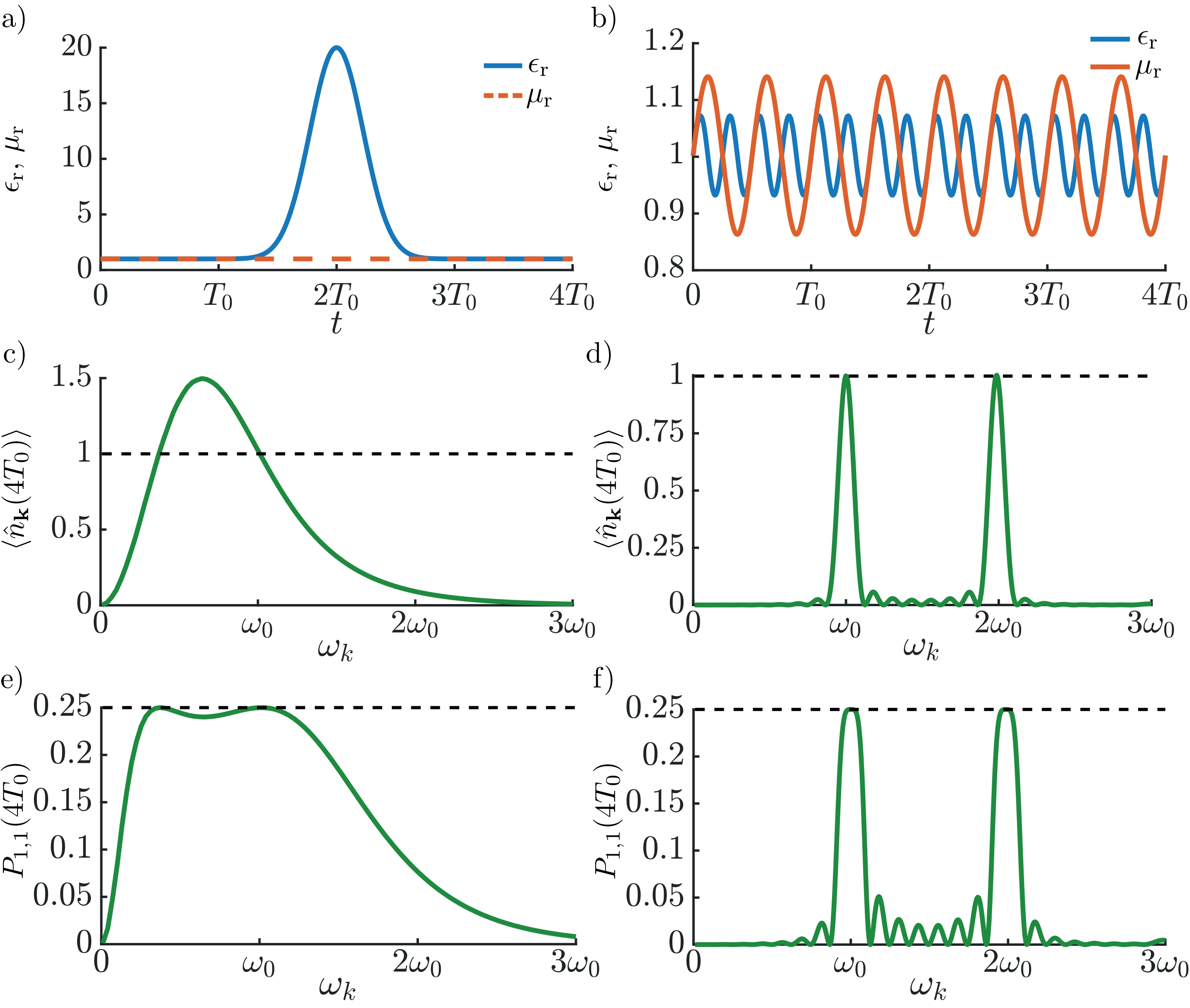}
    \caption{Emission spectra for a) Gaussian modulated permittivity and constant permeability, b) Sinusoidal permittivity and permeability modulation of frequency $4\omega_0$ and $2\omega_0$, respectively. c), d) Average photon number at time $t_0 = 4T_0$~[\eqref{Eq.: n(t_0)}] for the Gaussian modulation and the sinusoidal modulation, respectively. e), f) Probability of generating a single photon pair at time $t_0 = 4T_0$~[\eqref{Eq.: P(t_0)}] due to the Gaussian modulation and the sinusoidal modulation, respectively.}
    \label{Fig.: Spectral Profiles}
\end{figure}

The effect of the modulation profile becomes apparent when considering the spectral profile of the photons created from vacuum.
The differential equations that govern $f_k$ and $g_k$ in \eqref{Eq.: Differential equations f and g} depend on the photon frequency, $\omega_k$, 
through $\alpha_k(t)$ and $\beta_k(t)$.
Figure \ref{Fig.: Spectral Profiles} compares the spectral profile for a Gaussian permittivity modulation with the spectral profile for a sinusoidal permittivity and permeability modulation. Since we assume a non-dispersive medium, these profiles are only valid in frequency regions sufficiently far from material resonances, where the non-dispersive approximation remains accurate.
The average photon number at a specific time, $t_0$, can be calculated by substituting \eqref{Eq.: Ansatz} into 
\begin{equation}\label{Eq.: n(t_0)}
   \begin{aligned}
        \langle \hat{n}_{k}(t_0) \rangle = \bra{0, 0} \hat{a}^\dagger_{k}(t_0)\hat{a}_{k}(t_0) \ket{0, 0} = |g_k(t_0)|^2.
   \end{aligned}
\end{equation}
The Gaussian permittivity modulation results in a single, broad peak in the spectral profile, representing the average number of emitted photons.
In comparison, the sinusoidal permittivity and permeability modulations lead to two peaks in the spectral profile, 
indicating that there would on average be a single photon pair with frequency $\omega_0$ and a single photon pair with frequency $2\omega_0$ at time $t_0 = 4T_0$.
Similarly, we see that the probability of creating a single photon pair as a function of the photon's frequency is influenced by the modulation profile.
The single photon pair production probability is maximal when $\langle \hat{n}_{k}(4T_0) \rangle = 1$, so we two peaks appear in this probability for the sinusoidal modulation.
The Gaussian modulation considered here leads to average photon numbers higher than $1$ for certain frequencies, and we see that for these frequencies, $P_{1, 1}(4T_0)$ is lower than $25 \%$.

The instantaneous eigenstate method can also be applied to other initial states, such as the coherent state $\ket{\alpha, 0} = e^{-|\alpha|^2/2}\sum_n \frac{\alpha^n}{\sqrt{n!}} \ket{n, 0}$.
This state, which represents a plane wave, evolves in the time-varying medium into a state with coefficients $C_{n, m}(t)$ given by
\begin{equation}\label{Eq.: Coherent State Coefficients}
    \begin{aligned}
         C_{n,m}(t) = \sqrt{\frac{n!}{m!}} \frac{\alpha^{n-m}e^{-|\alpha|^2/2}g^m_k(t)}{(n-m)!(f^{n+1}_k(t))^*}\Theta(n-m),
    \end{aligned}
\end{equation}
which can be determined very similarly to what was done in Sec.~$5$ of \cite{Supplementary_material}.
While a classical plane wave traveling through a time-varying medium splits into a later-forward and later-backward wave \cite{morgenthaler_velocity_1958}, 
\eqref{Eq.: Coherent State Coefficients} shows that the forward and backward modes are actually entangled.
For example, the Heaviside step function, $\Theta(n-m)$, implies that there must be at least as many photons in the forward mode as in the backward mode.
Therefore, if a detector would measure a single photon in the forward mode, the backward mode would collapse into a state with either $0$ or $1$ photon.
This shows that the effect of time-varying media on light can be fully understood only by considering the quantum nature of light, including the creation and annihilation events that occur in the photon field.

\begin{figure}[!ht]
    \centering
    \includegraphics[width=\linewidth]{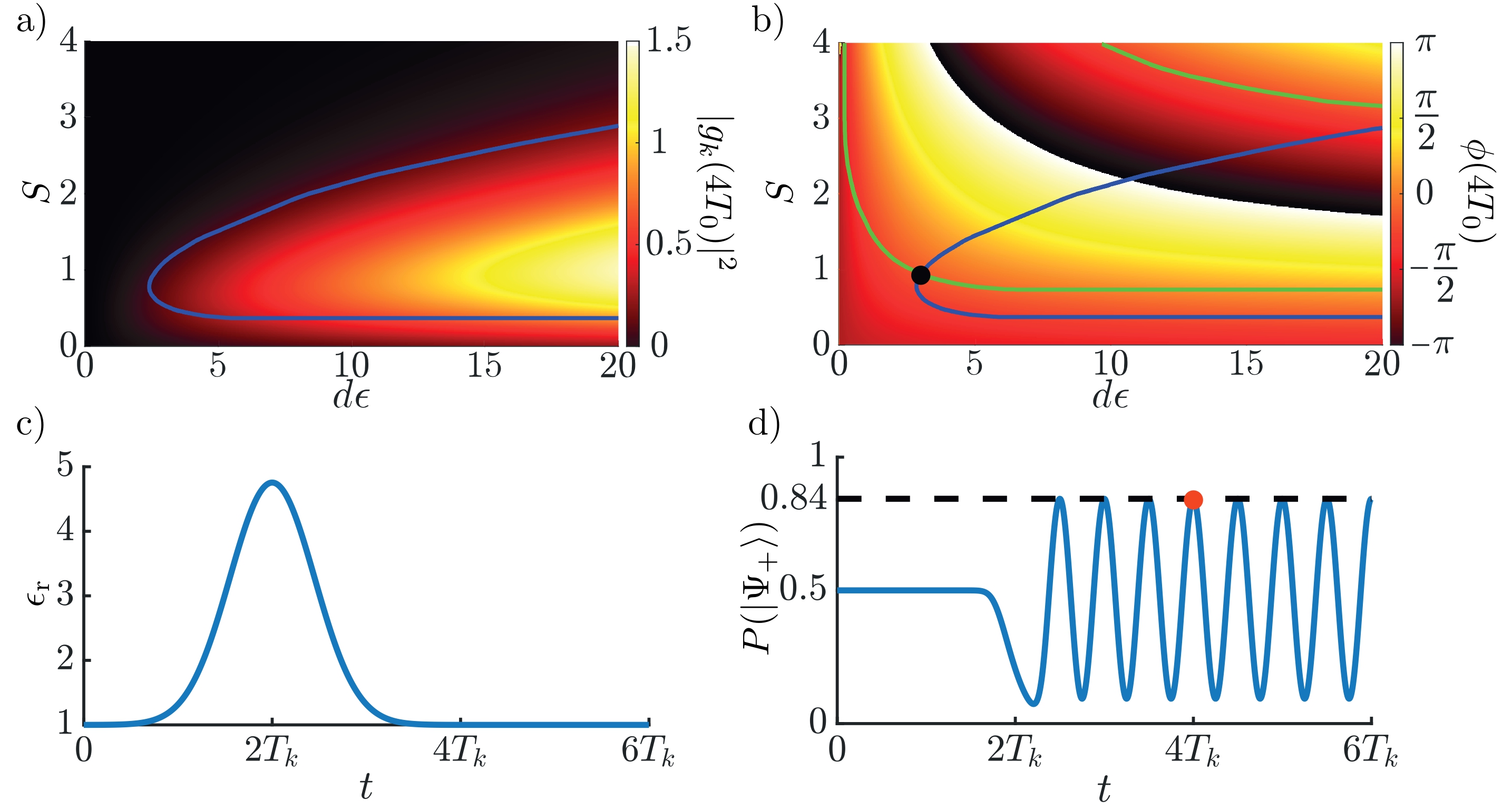}
    \caption{a), b) Value of $|g_k(4T_k)|$ and $\phi(4T_k)$ in a medium undergoing a Gaussian permittivity modulation $\epsilon_\rm{r}(t) = d\epsilon e^{-t^2/S^2}$ as a function of $d\epsilon$ and $S$ along with the level lines for $|g_k(4T_k)|^2 = 1/3$ and $\phi(4T_k) = 0$.
    c) Gaussian permittivity modulation $\epsilon_\rm{r}(t) = 3.75 e^{-t^2/1.16^2}$ which leads to $P(\ket{\Psi_+}) = 0.84$ at time $4T_k$. d) Evolution of the probability $P(\ket{\Psi_+})$~[\eqref{Eq.= P_Bell}] due to the permittivity modulation shown in~c), the red dot corresponds to the time $4T_k$, where $P(\ket{\Psi_+})=0.84$.}
    \label{Fig.: Gaussian Optimization}
\end{figure}

The instantaneous eigenstate method also allows us to calculate the probability of generating a desired state $\ket{\phi}$ using certain modulation profiles. This probability is given by $P(\ket{\phi}, t) = |\bra{\phi} \ket{\psi(t)}|^2$, where $\ket{\psi(0)}$ is the initial state of the system.
For example, to study the probability of generating the Bell state $\ket{\Psi_+} = (\ket{0,0} + \ket{1, 1})/\sqrt{2}$ from vacuum, $P(\ket{\Psi_+}, t) = |\bra{\Psi_+} \ket{\psi(t)}|^2$, is given, using \eqref{Eq.: Vacuum Evolution}, by
\begin{equation}\label{Eq.= P_Bell}
   \begin{aligned}
        P\left(\ket{\Psi_+}\right) = \frac{\left(1 + 2\frac{|g_k(t)|\cos \phi(t)}{|f_k(t)|} + \frac{|g_k(t)|^2}{|f_k(t)|^2} \right)}{2|f_k(t)|^2},
   \end{aligned}
\end{equation}
where $\phi(t) = \theta_f(t) + \theta_g(t)$ with $\theta_f(t)$ and $\theta_g(t)$ being the phases of $f_k(t)$ and $g_k(t)$, respectively.
From this formula, we can derive the maximal probability of generating  $\ket{\Psi_+}$.
This probability is $27/32 \approx 0.84$, and it occurs at $|g_k(t)|^2 = 1/3$ and $\phi = 0$.
We now know the maximum probability of generating $\ket{\Psi_+}$, but it remains unclear what type of modulation would simultaneously lead to $|g_k(t)|^2 = 1/3$ and $\phi = 0$ to achieve that maximum. Suppose that the initial state of the system, at $t=0$, is vacuum, $\ket{0, 0}$, and our goal is to generate a Bell state at a later time, say $4T_k$, where $T_k = 2\pi/\omega_k$ is the period of the photons that are created. One possible modulation profile is a Gaussian permittivity modulation $\epsilon_\rm{r}(t) = \epsilon(t)/\epsilon(0) = 1+ d\epsilon e^{-t^2/S^2}$. Figure \ref{Fig.: Gaussian Optimization} shows $|g_k(4 T_k)|^2$ and $\phi(4 T_k)$ as functions of $d\epsilon$ and $S$, which allows us to identify suitable values for the parameters $d\epsilon$ and $S$ to maximize the probability of generating the Bell state.
The modulation with these optimal parameters, along with the evolution of $P(\ket{\Psi_+})$, is shown in \figref{Fig.: Gaussian Optimization} as well.
Notice that the probability continues to oscillate after reaching its maximum of $84\%$ at time $4T_k$, even though the modulation has ended at that time. 
This is because $f_k(t)$ and $g_k(t)$ maintain their oscillating phase factor.
When $\epsilon_\rm{r}(t) = 1$ and $\mu_\rm{r}(t) = 1$, the coefficient $\beta_k(t) \propto \frac{\epsilon_1}{\epsilon(t)} - \frac{\mu_1}{\mu(t)}$ vanishes and $\alpha_k(t) = \omega_k$.
Then \eqref{Eq.: Differential equations f and g} implies that $f_k(t)$ and $g_k(t)$ oscillate with the same frequency as the created photons, $\omega_k$.
Therefore, the phase factor $\phi(t) = \theta_f(t) + \theta_g(t)$ in \eqref{Eq.= P_Bell} causes $P\left(\ket{\Psi_+}\right)$ to continue oscillating after the modulation of the permittivity has ended.
This behavior is due to the fact that an initial state $\ket{\psi(0)} = \ket{\Psi_+}$ in an unmodulated medium evolves as $\ket{\psi(t)} = (e^{-i \omega_k t} \ket{0, 0} + e^{-3i \omega_k t} \ket{1, 1})/\sqrt{2}$, as may be verified by solving the Schrödinger equation with $\hat{H}(t) = \hbar \omega_k (\hat{n}_\rm{k} + \hat{n}_\rm{-k} + 1)$, which indicates an oscillation between $\ket{\Psi_+}$ and $\ket{\Psi_-}=(\ket{0, 0} - \ket{1, 1})/\sqrt{2})$. 
Therefore, when we produce a Bell state $\ket{\Psi_+}$ with $84\%$, the state will still oscillate between $\ket{\Psi_+}$ and $\ket{\Psi_-}$, causing the oscillation in $P(\ket{\Psi_+})$ after the modulation has ended.

\begin{figure}[!ht]
    \centering
    \includegraphics[width=\linewidth]{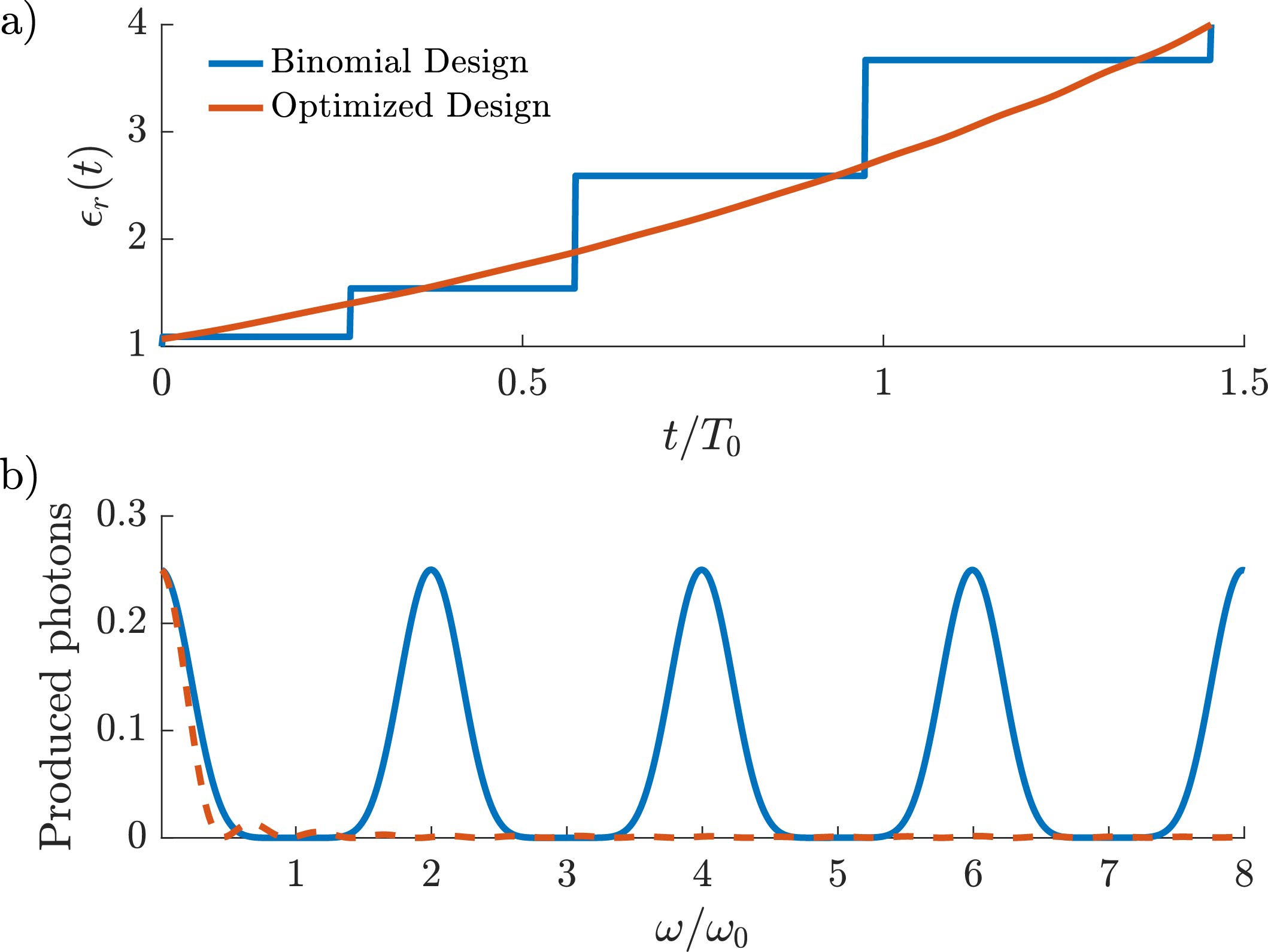}
    \caption{a) Permittivity modulation for the binomial-design ATC with discrete permittivity steps from $\epsilon_r$ = 1.09 at $t=0$, $\epsilon_r$ = 1.54 at $t=0.261T_0$, $\epsilon_r$ = 2.59 at $t=0.572T_0$, $\epsilon_r$ = 3.67 at $t=0.975T_0$, $\epsilon_r$ = 4 at $t=1.454T_0$ and for the optimized ATC described in \cite{Supplementary_material}. b) Generated noisy photons for the binomial-design and optimized ATC as a function of the frequency $\omega$, assuming the signal consists out of a single photon with frequency $\omega$.}
    \label{Fig.: ATC}
\end{figure}

The instantaneous eigenstate method allows us not only to calculate how the electromagnetic field responds to variations in the constitutive parameters, but also to design these parameter variations in order to shape the electromagnetic field itself.
To illustrate this idea, we we show how a carefully engineered permittivity modulation can enhance the performance of quantum antireflection temporal coatings (ATCs) \cite{pacheco-pena_antireflection_2020, liberal_quantum_2023, castaldi2021exploiting}.
In an ATC, the refractive index of a medium is varied in time so that the frequency of an incoming signal, $\omega_1$, is converted to $\omega_2 = \frac{n_2}{n_1}\omega_1$.
Generally, this temporal change generates unwanted photons which manifest as noise on the converted signal. However, by selecting an appropriate temporal modulation of the refractive index, the photons created during the process can be made to mutually annihilate by the end of the operation, ensuring that the signal amplitude—or equivalently the photon number—remains unchanged.
In \figref{Fig.: ATC}, the blue curve shows a stepwise permittivity modulation designed to minimize photon generation around a target frequency $\omega_0$, following the binomial design of \cite{castaldi2021exploiting}. Such stepwise profiles, however, inevitably introduce periodic peaks in the noise spectrum, including at high frequencies.
In Sec. 6 of \cite{Supplementary_material}, we explain how to compute a continuous permittivity profile that suppresses ATC-induced noise over an arbitrary frequency band. The orange curve in \figref{Fig.: ATC} shows such a profile, optimized to minimize noise for frequencies $\omega \in [0.01\omega_0, 6\omega_0]$, with permittivity variations limited to $\epsilon_\rm{r}(t)\in [0.1, 5]$ and a total operation time of $1.454T_0$, which coincides with the operation time of the binomial design.
This optimized modulation slightly reduces the noise at frequencies far below $\omega_0$, slightly increases it near $\omega_0$, and produces a substantial reduction at higher frequencies, even beyond $6\omega_0$.
This strong high-frequency suppression arises from the continuity of the calculated permittivity profile: high-frequency components experience an effectively adiabatic change of the permittivity and therefore do not generate photons.
This example highlights only one way in which continuous modulations can be exploited to design ATCs. One could similarly target noise suppression around frequencies closer to $\omega_0$, or reduce the operational time of the ATC while still achieving acceptable noise suppression, enabling ultrafast frequency conversion.

We have introduced the instantaneous eigenstate method to study the evolution of an arbitrary initial state in a medium with arbitrary time-varying permittivity and permeability.
This method, grounded in the Heisenberg equation, allows to calculate the state evolution by solving only two coupled differential equations, while the Schrödinger equation would require solving an infinite set of coupled differential equations.
The method also provides general insights into the photon statistics of time-varying media.
We have established that the probability of generating a single photon pair is limited to $25\%$ for any modulation profile and that, similarly, the probability of generating the Bell state $\ket{\Psi_+}$ is limited to $27/32 \approx 84\%$.
Furthermore, we have shown that these maximum probabilities can be achieved by selecting an appropriate modulation profile.
Finally, we demonstrated how a temporal modulation profile enables affects the emission spectrum of generated photons and how we can determine the modulation profile providing a specified spectrum, thereby offering new possibilities for tailoring quantum light such as broadband ATC design for photon frequency conversion.

In future research, the instantaneous eigenstate method could be extended to analyze systems with space- and time-varying permittivity and permeability modulations. Furthermore, this method could be employed to investigate systems where a forward mode is coupled not only to a backward mode but also to additional modes, offering broader applicability to more complex interactions.
In this work, we have neglected dispersion, a valid approximation in frequency ranges far from material resonances. However, assessing the impact of dispersion on photon statistics and the emission spectrum represents an important direction for future research~\cite{sloan_casimir_2021, sloan_optical_2024, horsley_macroscopic_2024, koutserimpas_time-varying_2024}.



\newpage
\bibliography{bibliografie}

\newpage
\onecolumngrid
\renewcommand{\appendixname}{Section}
\newpage
\section{Supplementary Material}
\setcounter{equation}{0} 
\renewcommand{\theequation}{S\arabic{equation}} 
\input{SupMat}

\end{document}

%% file: SupMat.tex
\fontsize{10pt}{8pt}\selectfont

\setcounter{secnumdepth}{1}
\renewcommand{\thesection}{\arabic{section}}

\section{Time-Dependent Hamiltonian [Eq.~(4)]}\label{Sec.: Time-Dependent Hamiltonian}
The Hamiltonian of a system is the operator that leads to the appropriate equations of motion.
In a system with time-varying permittivity and permeability, the equations of motion are the Maxwell's equations, which may be written as
\begin{equation}
   \begin{aligned}
      \epsilon(t) \div{\b{E}} &= 0,  &\qquad \qquad \curl{\b{E}} &= - \frac{\partial \b{B}}{\partial t},\\ 
      \div{\b{B}} &= 0,  &\qquad \qquad \curl{\b{B}} &= \mu(t)\frac{\partial \b{D}}{\partial t}.
   \end{aligned}
\end{equation}
However, if we use the vector potential $\b{A}$ and scalar potential $\phi$ defined through $\b{E} = -\partial_t \b{A} - \nabla \phi$ and $\b{B} = \curl{\b{A}}$, and adopt the radiation gauge, 
$\div{\b{A}} = 0$ and $\phi = 0$, three of these equations are automatically satisfied. Indeed, the electric field equations simplify to $\b{E} = -\partial_t \b{A}$ and $\div{\b{E}} = 0$.
Next, the divergence of a curl is always zero, so that $\div{(\curl{\b{A}})} = 0$, which implies $\div{\b{B}} = 0$.
Finally, the curl of the electric field satisfies $\curl{\b{E}} = -\partial_t (\curl{\b{A}}) = - \partial_t \b{B}$. 
Thus, there is only only equation that is not fulfilled yet in the radiation gauge: $\curl{\b{B}} = \mu(t)\frac{\partial \b{D}}{\partial t}$. 
Therefore, the Hamiltonian of the system should lead to this equation.
\par
We will now prove that the Hamiltonian density
\begin{equation}\label{Eq.: Hamiltonian Density}
   \begin{aligned}
        \mathcal{H}(t) &= \frac{1}{2} \left(  \frac{\b{D}^2(\b{r}, t)}{\epsilon(t)}   +  \frac{(\curl{\b{A}}(\b{r}, t))^2}{\mu(t)}   \right)
   \end{aligned}
\end{equation}
will indeed lead to the final Maxwell equation.
In the Hamiltonian field theory for linear media, the fields are given by the vector potential components $A_n$ and the conjugate fields are the electric displacement components $-D_n$~\cite{hillery_introduction_2009}.
The equations of motion for these fields are given by \cite{goldstein_classical_2002}
\begin{equation}\label{Eq.: Hamiltonian Field Theory}
   \left\{
       \begin{aligned}
            \partial_t A_n &= \frac{\partial \mathcal{H}}{\partial (-D_n)}\\ 
            \partial_t (-D_n) &= -\frac{\partial \mathcal{H}}{\partial A_n} + \partial_i \frac{\partial \mathcal{H}}{\partial (\partial_i A_n)}
       \end{aligned}
   \right.
\end{equation}
The first equation in \eqref{Eq.: Hamiltonian Field Theory} results in $\partial_t A_n = -D_n/\epsilon(t) = -E_n$ which is consistent with the radiation gauge.
For the second equation in \eqref{Eq.: Hamiltonian Field Theory}, $-\frac{\partial \mathcal{H}}{\partial A_n} = 0$ since the Hamiltonian density does not depend explicitly on $\b{A}$.
Next, we can rewrite the curl of the vector potential as $\curl{\b{A}} = \varepsilon^{\mu \nu \sigma} \hat{e}_\mu \partial_\nu A_\sigma$, where 
$\varepsilon^{\mu \nu \sigma}$ is the Levi-Civita symbol and $\hat{e}_\mu$ are the unit vectors in the $x$, $y$ and $z$ directions.
Then, $\frac{\partial \curl{\b{A}}}{\partial (\partial_i A_n)} = \varepsilon^{\mu i n} \hat{e}_\mu$ such that 
$\frac{\partial \mathcal{H}}{\partial (\partial_i A_n)} = \frac{\epsilon^{\mu i n}}{\mu(t)} \b{B} \cdot  \hat{e}_\mu = \frac{1}{\mu(t)} \epsilon^{\mu i n} \b{B}_\mu$.
It follows that $\partial_i \frac{\partial \mathcal{H}}{\partial (\partial_i A_n)} = \frac{1}{\mu(t)} \epsilon^{\mu i n} \partial_i \b{B}_\mu$.
When we reorder the indices in the Levi-Civita symbol, we find $\epsilon^{\mu i n} \partial_i \b{B}_\mu = -(\curl{\b{B}})_n$.
Finally, we conclude that the second equation in \eqref{Eq.: Hamiltonian Field Theory} results in $\curl{\b{B}} = \mu(t) \partial_t \b{D}$, which validates the Hamiltonian density in \eqref{Eq.: Hamiltonian Density}.

To express the Hamiltonian in terms of creation and annihilation operators, we substitute 
\begin{equation}\label{Eq.:}
   \begin{aligned}
    \hat{\b{A}}(\b{r}, t)  &=\sum_{\b{k}, \lambda} \sqrt{\frac{\hbar}{2\epsilon_1 \omega_k  V}} \b{e}_{\b{k},\lambda} \left( \hat{a}_{\b{k}, \lambda}(t)e^{i\b{k}\cdot\b{r}} +  \hat{a}^\dagger_{\b{k}, \lambda}(t)e^{-i\b{k}\cdot\b{r}}\right), \\
    \hat{\b{D}}(\b{r}, t)  &=\sum_{\b{k}, \lambda}i \sqrt{\frac{\hbar \epsilon_1 \omega_k}{2V}} \b{e}_{\b{k},\lambda} \left(   \hat{a}_{\b{k}, \lambda}(t) e^{i\b{k}\cdot\b{r}}-  \hat{a}^\dagger_{\b{k}, \lambda}(t)e^{-i\b{k}\cdot\b{r}}\right),\\    
   \end{aligned}
\end{equation}
into the Hamiltonian density \eqref{Eq.: Hamiltonian Density} and integrate over the volume of the cavity $V$.
The curl of the vector potential, $\hat{A}$, is
\begin{equation}
   \begin{aligned}
      \curl{\hat{\b{A}}(\b{r}, t)}  &=\sum_{\b{k}, \lambda} i \sqrt{\frac{\hbar}{2\epsilon_1 \omega_k  V}} (\b{k} \times \b{e}_{\b{k},\lambda}) \left( \hat{a}_{\b{k}, \lambda}(t)e^{i\b{k} \cdot \b{r}} -  \hat{a}^\dagger_{\b{k}, \lambda}(t)e^{-i\b{k} \cdot \b{r}}\right), \\
   \end{aligned}
\end{equation}
and the square of this expression gives 
\begin{equation}\label{Eq.: curl^2}
   \begin{aligned}
      \hspace{-5mm} \left(\curl{\hat{\b{A}}(\b{r}, t)}\right)^2 &= \sum_{\b{k}, \b{q}, \lambda, \nu} \frac{-\hbar(\b{k} \times \b{e}_{\b{k},\lambda}) \cdot (\b{q} \times \b{e}_{\b{q},\nu})}{2\epsilon_1 V \sqrt{\omega_k \omega_q}}  
      \left( \hat{a}_{\b{k}, \lambda}(t) \hat{a}_{\b{q}, \nu}(t)e^{i(\b{k} + \b{q}) \cdot \b{r}} +  \hat{a}^\dagger_{\b{k}, \lambda}(t) \hat{a}^\dagger_{\b{q}, \nu}(t)e^{-i(\b{k} + \b{q}) \cdot \b{r}}\right)\\ 
      &\quad + \sum_{\b{k}, \b{q}, \lambda, \nu} \frac{\hbar(\b{k} \times \b{e}_{\b{k},\lambda}) \cdot (\b{q} \times \b{e}_{\b{q},\nu})}{2\epsilon_1 V \sqrt{\omega_k \omega_q}}  
      \left( \hat{a}_{\b{k}, \lambda}(t) \hat{a}^\dagger_{\b{q}, \nu}(t)e^{i(\b{k} - \b{q}) \cdot \b{r}} +  \hat{a}^\dagger_{\b{k}, \lambda}(t) \hat{a}_{\b{q}, \nu}(t)e^{-i(\b{k} - \b{q}) \cdot \b{r}}\right).\\ 
   \end{aligned}
\end{equation}
Since the wave vectors are quantized as $\b{k} = \frac{2\pi}{L}(n_x, n_y, n_z)$, where $n_x$, $n_y$ and $n_z$ are integers, we find $\int_V d^3 r e^{i(\b{k}-\b{q}) \cdot \b{r}} = V\delta_{\b{k}, \b{q}}$.
Integrating \eqref{Eq.: curl^2} over space then results in
\begin{equation}\label{Eq.: curl^2 integral}
   \begin{aligned}
      \int_V  \rm{d}^3 r \left(\curl{\hat{\b{A}}(\b{r}, t)}\right)^2 &= \sum_{\b{k}, \lambda, \nu} \frac{-\hbar(\b{k} \times \b{e}_{\b{k},\lambda}) \cdot (-\b{k} \times \b{e}_{-\b{k},\nu})}{2\epsilon_1 \sqrt{\omega_k \omega_{-k}}}  
      \left( \hat{a}_{\b{k}, \lambda}(t) \hat{a}_{-\b{k}, \nu}(t) +  \hat{a}^\dagger_{\b{k}, \lambda}(t) \hat{a}^\dagger_{-\b{k}, \nu}(t)\right)\\ 
      &\quad + \sum_{\b{k}, \lambda, \nu} \frac{\hbar(\b{k} \times \b{e}_{\b{k},\lambda}) \cdot (\b{k} \times \b{e}_{\b{k},\nu})}{2\epsilon_1 \sqrt{\omega_k \omega_k}}  
      \left( \hat{a}_{\b{k}, \lambda}(t) \hat{a}^\dagger_{\b{k}, \nu}(t) +  \hat{a}^\dagger_{\b{k}, \lambda}(t) \hat{a}_{\b{k}, \nu}(t)\right).\\ 
   \end{aligned}
\end{equation}
Since the wave vector $\b{k}$ and the polarization vectors $\b{e}_{\b{k}, 1}$ and $\b{e}_{\b{k}, 2}$ consitute an orthonormal basis, where we can choose 
$\b{e}_{-\b{k},\lambda} = (-1)^\lambda \b{e}_{\b{k},\lambda}$~\cite{hillery_introduction_2009}, the Cauchy-Binet identity, $(\b{a} \times \b{b}) \cdot (\b{c} \times \b{d}) = (\b{a} \cdot \b{c})(\b{b} \cdot \b{d}) - (\b{a} \cdot \b{d})(\b{b} \cdot \b{c})$,
implies  
\begin{equation}
   \left\{
   \begin{aligned}
      (\b{k} \times \b{e}_{\b{k},\lambda}) \cdot (-\b{k} \times \b{e}_{-\b{k},\nu}) &= (-1)^\lambda k^2 \delta_{\mu, \nu},\\ 
      (\b{k} \times \b{e}_{\b{k},\lambda}) \cdot (\b{k} \times \b{e}_{\b{k},\nu}) &= k^2 \delta_{\lambda, \nu}.
   \end{aligned}
   \right.
\end{equation}
We substitute these equations into \eqref{Eq.: curl^2 integral}, along with $\omega_{-k} = \omega_k$, to find 
\begin{equation}\label{Eq.: curl^2 integral 2}
   \begin{aligned}
      \int_V  \rm{d}^3 r \left(\curl{\hat{\b{A}}(\b{r}, t)}\right)^2 &= \sum_{\b{k}, \lambda} \frac{(-1)^{\lambda+1}\hbar k^2}{2\epsilon_1 \omega_k }  
      \left( \hat{a}_{\b{k}, \lambda}(t) \hat{a}_{-\b{k}, \lambda}(t) +  \hat{a}^\dagger_{\b{k}, \lambda}(t) \hat{a}^\dagger_{-\b{k}, \lambda}(t)\right)\\ 
      &\quad + \sum_{\b{k}, \lambda} \frac{\hbar k^2}{2\epsilon_1 \omega_k}  
      \left( \hat{a}_{\b{k}, \lambda}(t) \hat{a}^\dagger_{\b{k}, \lambda}(t) +  \hat{a}^\dagger_{\b{k}, \lambda}(t) \hat{a}_{\b{k}, \lambda}(t)\right).\\ 
   \end{aligned}
\end{equation}
The calculation of $\int_V \rm{d}^3 r \b{D}^2(\b{r}, t)$ is similar and results into
\begin{equation}\label{Eq.: integral D}
   \begin{aligned}
      \int_V \rm{d}^3 r \b{D}^2(\b{r}, t) &= \sum_{\b{k}, \lambda} \frac{(-1)^{\lambda} \hbar \epsilon_1 \omega_k}{2}    
      \left( \hat{a}_{\b{k}, \lambda}(t) \hat{a}_{-\b{k}, \lambda}(t) +  \hat{a}^\dagger_{\b{k}, \lambda}(t) \hat{a}^\dagger_{-\b{k}, \lambda}(t)\right)\\ 
      &\quad + \sum_{\b{k}, \lambda} \frac{\hbar \epsilon_1 \omega_k}{2}  
      \left( \hat{a}_{\b{k}, \lambda}(t) \hat{a}^\dagger_{\b{k}, \lambda}(t) +  \hat{a}^\dagger_{\b{k}, \lambda}(t) \hat{a}_{\b{k}, \lambda}(t)\right).\\ 
   \end{aligned}
\end{equation}
Because of the commutation relation $[\hat{a}_{\b{k}}(t), \hat{a}^\dagger_{\b{k}}(t)] = 1$, we have $\hat{a}_{\b{k}, \lambda}(t) \hat{a}^\dagger_{\b{k}, \lambda}(t) +  \hat{a}^\dagger_{\b{k}, \lambda}(t) \hat{a}_{\b{k}, \lambda}(t) = 2 \hat{n}_{\b{k}, \lambda}(t) + 1$, 
and thus with \eqref{Eq.: Hamiltonian Density}, \eqref{Eq.: curl^2 integral 2} and \eqref{Eq.: integral D}, the Hamiltonian becomes 
\begin{equation}\label{Eq.: H(a, ad, t)S}
   \begin{aligned}
   \hat{H}(t) = \int_V \rm{d}^3 r \mathcal{H}(t) &= \sum_{\bf{k}, \lambda} \hbar \alpha_k(t) \left(   \hat{n}_{\bf{k}, \lambda}(t) + \frac{1}{2}     \right)
   + \frac{\hbar \beta_k(t)}{2} \left(\hat{a}_{\bf{k}, \lambda}(t)  \hat{a}_{-\bf{k}, \lambda}(t)
   +
   \hat{a}^\dagger_{\bf{k}, \lambda}(t) \hat{a}^\dagger_{-\bf{k}, \lambda}(t) \right),\\
   \end{aligned}
\end{equation}
with $\alpha_k(t) = \frac{\omega_k}{2}\left(\frac{\epsilon_1}{\epsilon(t)} + \frac{\mu_1}{\mu(t)}\right)$ and $\beta_k(t) = \frac{(-1)^\lambda \omega_k}{2}\left(\frac{\epsilon_1}{\epsilon(t)} - \frac{\mu_1}{\mu(t)}\right)$.

\newpage

\section{Schrödinger Equation}\label{Sec.: Schrödinger Equation}
We consider the photon state evolution in a single forward mode ($\b{k}, \lambda$) and related backward mode ($-\b{k}, \lambda$).
The state of the photon field is then $\ket{\psi(t)} = \sum_{n, m=0}^\infty C_{n, m}(t) \ket{n, m}$, where $\ket{n, m} \equiv \ket{n}_{\b{k}, \lambda} \ket{m}_{-\b{k}, \lambda}$.
To determine the evolution of this state, we need to calculate the coefficients $C_{n, m}(t)$, which are governed by the Schrödinger equation $i \hbar \partial_t \ket{\psi(t)} = \hat{H}(t) \ket{\psi(t)}$.
Since we are considering only the modes ($\b{k}, \lambda$) and ($-\b{k}, \lambda$), the Hamiltonian can be restricted to these modes, so that \eqref{Eq.: H(a, ad, t)S} reduces to
\begin{equation}
   \begin{aligned}
    \hat{H}(t) &= \hbar \alpha_k(t) \left(   \hat{n}_{\bf{k}, \lambda} + \hat{n}_{-\bf{k}, \lambda} + 1     \right) 
    + \hbar \beta_k(t) \left(\hat{a}_{\bf{k}, \lambda}  \hat{a}_{-\bf{k}, \lambda}
     +
    \hat{a}^\dagger_{\bf{k}, \lambda} \hat{a}^\dagger_{-\bf{k}, \lambda} \right).\\
   \end{aligned}
\end{equation}
When we use the operator relations $\hat{a}^\dagger_{\bf{k}, \lambda} \ket{n, m} = \sqrt{n+1}\ket{n+1, m}$, $\hat{a}_{\bf{k}, \lambda} \ket{n, m} = \sqrt{n}\ket{n-1, m}$ 
and $\hat{n}_{\bf{k}, \lambda} \ket{n, m} = n \ket{n, m}$, we find that the right-hand side of the Schrödinger equation equals
\begin{equation}
   \begin{aligned}
        \hat{H}(t)\ket{\psi(t)} &= \hbar \alpha_k(t) \sum_{n, m} C_{n, m}(t) \left( n + m + 1 \right) \ket{n, m} \\
        &\quad + \hbar \beta_k(t) \sum_{n=1, m=1} C_{n, m}(t) \sqrt{nm} \ket{n-1, m-1} \\ 
        &\quad + \hbar \beta_k(t) \sum_{n, m} C_{n, m}(t) \sqrt{(n+1)(m+1)} \ket{n+1, m+1}.
   \end{aligned}
\end{equation}
On the other hand, the left-hand side of the Schrödinger equation is $\partial_t \ket{\psi(t)} = \sum_{n,m} \partial_t C_{n,m}(t) \ket{n, m}$.
If we now act upon both sides of the Schrödinger equation with $\bra{n', m'}$ and use the orthonormality of Fock states, $\bra{n', m'}\ket{n, m} = \delta_{n, n'}\delta_{m,m'}$,
we find the differential equations for the coefficients $C_{n', m'}(t)$,
 \begin{equation}\label{Eq.: Schrödinger Equation}
    \begin{aligned}
        i \partial_t C_{n',m'}(t) &= \alpha_k(t) \left( n'+ m' + 1 \right) C_{n', m'}(t) + \beta_k(t) \sqrt{(n'+1)(m'+1)} C_{n'+1, m'+1}(t) \\ 
           &\quad +  \beta_k(t) \sqrt{n'm'} C_{n'-1, m'-1}(t),
    \end{aligned}
 \end{equation}
where we define $C_{n, m}(t) = 0$ for $n < 0$ or $m < 0$.

Equation~(\ref{Eq.: Schrödinger Equation}) actually represents an infinite set ($n',m'=0\ldots\infty$) of coupled differential equations with time-varying coefficients. This is both analytically difficult and computationally intensive. Moreover, exploring different temporal profiles would require solving these equations repeatedly, which would drastically increase the required computational resources. Therefore, we require a different approach to gain a deeper understanding of how temporal modulation profiles affect the photon field.

\newpage
\section{Instantaneous Ground State [Eq.~(13)]}\label{Sec.: Instantaneous Ground State}

In the Heisenberg picture, the evolution of an operator $\hat{A}(0)$ is given by $\hat{A}(t) = \hat{U}^\dagger(t) \hat{A}(0) \hat{U}(t)$, where $\hat{U}(t) = e^{-i\hat{H}t/\hbar}$ is the unitary time-evolution operator.
This unitarity, along with $\ket{\xi_{n, m}(t)} = \hat{U}^\dagger(t) \ket{n, m}$, implies
\begin{equation}
   \begin{aligned}
      \hat{A}(t) \ket{\xi_{n, m}(t)} = \hat{U}^\dagger(t) \hat{A}(0) \ket{n, m}.
   \end{aligned}
\end{equation}
In the case of the number operator $\hat{n}_\b{k}(0)$, we have $\hat{A}(0) = \hat{n}_\b{k}(0)$ and $\hat{n}_\b{k}(0) \ket{n, m} = n \ket{n, m}$, so that 
\begin{equation}
   \begin{aligned}
      \hat{n}_\b{k}(t) \ket{\xi_{n, m}(t)} = \hat{U}^\dagger(t) n \ket{n, m}.
   \end{aligned}
\end{equation}
Since $n$ is a scalar, it commutes with the time-evolution operator and using again $\ket{\xi_{n, m}(t)} = \hat{U}^\dagger(t) \ket{n, m}$, we find 
\begin{equation}
   \begin{aligned}
      \hat{n}_\b{k}(t) \ket{\xi_{n, m}(t)} = n \ket{\xi_{n, m}(t)}.
   \end{aligned}
\end{equation}
Therefore, the states $\ket{\xi_{n, m}(t)}$ are the instantaneous eigenstates of the number operator $\hat{n}_\b{k}(t)$.
Similarly, $\hat{a}_\b{k}(0) \ket{n, m} = \sqrt{n} \ket{n-1, m}$ and $\hat{a}^\dagger_\b{k}(0) \ket{n, m} = \sqrt{n+1} \ket{n+1, m}$ imply 
\begin{equation}
   \begin{aligned}
      \hat{a}_\b{k}(t) \ket{\xi_{n, m}(t)} &= \hat{U}^\dagger(t) \sqrt{n} \ket{n-1, m} = \sqrt{n} \ket{\xi_{n-1, m}(t)},\\
      \hat{a}^\dagger_\b{k}(t) \ket{\xi_{n, m}(t)} &= \hat{U}^\dagger(t) \sqrt{n+1} \ket{n+1, m} = \sqrt{n+1} \ket{\xi_{n+1, m}(t)}.
\end{aligned}
\end{equation}
The instantaneous ground state $\ket{\xi_{0,0}(t)}$ can be expanded in the Fock state basis from before, so $\ket{\xi_{0,0}(t)} = \sum_{n, m=0}^\infty D_{n, m}(t) \ket{n, m}$.
Since the time-dependent annihilation operators $\hat{a}_{\pm \b{k}, \lambda}(t)$ lower an instantaneous eigenstate, and the instantaneous ground state is the lowest state, 
we have $\hat{a}_{\pm \b{k}, \lambda}(t) \ket{\xi_{0,0}(t)} = 0$.
With $\hat{a}_{\pm \b{k}, \lambda}(t) = f_k(t)\hat{a}_{\pm \b{k}, \lambda} + g_k(t)\hat{a}^\dagger_{\mp \b{k}, \lambda}$ and the expansion of $\ket{\xi_{0,0}(t)}$, these conditions become 
\begin{equation}
   \left\{
       \begin{aligned}
           f_k(t) \sum_{n=1, m=0} D_{n, m}(t) \sqrt{n}\ket{n-1, m} + g_k(t) \sum_{n=0, m=0} D_{n, m}(t) \sqrt{m+1}\ket{n, m+1} &= 0\\
           f_k(t) \sum_{n=0, m=1} D_{n, m}(t) \sqrt{m}\ket{n, m-1} + g_k(t) \sum_{n=0, m=0} D_{n, m}(t) \sqrt{n+1}\ket{n+1, m} &= 0\\
       \end{aligned}
   \right.
\end{equation}
We can then rewrite these sums to obtain relations between the coefficients $D_{n, m}(t)$,
\begin{equation}\label{Eq.: D_nm relations}
   \left\{
       \begin{aligned}
           D_{n+1, 0}(t) &= 0\\
            D_{0, m+1}(t) &= 0\\
            D_{n, m}(t) &= -\frac{g_k(t)\sqrt{m}}{f_k(t) \sqrt{n}}D_{n-1, m-1} \quad, n \geq 1\\ 
            D_{n, m}(t) &= -\frac{g_k(t)\sqrt{n}}{f_k(t) \sqrt{m}}D_{n-1, m-1} \quad, m \geq 1\\ 
       \end{aligned}
   \right.
\end{equation}
The two last equations can be combined into $n D_{n-1, m-1} = m D_{n-1, m-1}$, which implies that for $n\neq m$, $D_{n, m}(t) = 0$.
The only nonvanishing coefficients is then $D_{n, n}(t)$, and repeatedly using the last equation in \eqref{Eq.: D_nm relations} then results into $D_{n, n}(t) =  \left(-\frac{g_k(t)}{f_k(t)}\right)^n D_{0, 0}(t)$.
This coefficient $D_{0, 0}(t)$ can be determined by normalizing the instantaneous ground state.
Since $\braket{\xi_{0,0}(t)}{\xi_{0,0}(t)} = \sum_{n} \abs{D_{n, n}(t)}^2$ and $\sum_{n} \left(\frac{|g_k(t)|^2}{|f_k(t)|^2}\right)^{n} = |f_k(t)|^2$, using $|f_k(t)|^2 - |g_k(t)|^2=1$, which comes from the commutation relation $[\hat{a}_{\b{k}}(t), \hat{a}^\dagger_{\b{k}}(t)] =1$, 
we conclude that $|D_{0, 0}(t)|^2 = \frac{1}{|f_k(t)|^2}$.
This determines the instantaneous ground state up to a phase factor $\theta(t)$
\begin{equation}\label{Eq.: Instantaneous Ground State}
   \begin{aligned}
      \ket{\xi_{0,0}(t)} &= e^{i\theta(t)} \sum_{n} \frac{g^n_k(t)}{f_k^{n+1}(t)} \ket{n, n}
   \end{aligned}
\end{equation}

\newpage 

\section{Solutions to Heisenberg Equations for Temporal Step [Eq.~(17)]}
The differential equations 
\begin{equation}\label{Eq.: f, g}
   \left\{
       \begin{aligned}
        \partial_t f_k(t) &= -i\alpha_k(t)f_k(t) -i  \beta_k(t)g_k^*(t), \quad f_k(0) = 1,\\
        \partial_t g_k(t) &= -i\alpha_k(t)g_k(t) -i \beta_k(t)f^*_k(t), \quad g_k(0) = 0,\\
       \end{aligned}
   \right.
\end{equation}
are generally difficult to solve, but they can be analytically solved for temporal step modulations.
For a temporal step, the permittivity and permeability of the system change instantaneously at $t=0$ to $\epsilon_2$ and $\mu_2$ and then remain fixed.
Consequently, $\alpha_k(t) = \frac{\omega_k}{2}\left(\frac{\epsilon_1}{\epsilon(t)} + \frac{\mu_1}{\mu(t)}\right)$ and $\beta_k(t) = \frac{(-1)^\lambda \omega_k}{2}\left(\frac{\epsilon_1}{\epsilon(t)} - \frac{\mu_1}{\mu(t)}\right)$ become constant for $t > 0$, simplifying \eqref{Eq.: f, g} to
\begin{equation}
   \begin{aligned}
      \partial_t
      \begin{pmatrix}
         f_k(t)\\
         g_k^*(t)\\
      \end{pmatrix}
      = i\begin{pmatrix}
         -\alpha_k & -\beta_k\\ 
         \alpha_k & \beta_k\\
      \end{pmatrix}
      \begin{pmatrix}
         f_k(t)\\ 
         g_k^*(t)\\
      \end{pmatrix}.
   \end{aligned}
\end{equation}
We then diagonalize the coefficient matrix and define $\lambda_k = \frac{\omega_k}{\sqrt{\epsilon_{\rm{r}} \mu_{\rm{r}}}}$ so that
\begin{equation}\label{Eq.: diagonal diff.eq.}
   \begin{aligned}
      \partial_t
      \begin{pmatrix}
         f_k(t)\\
         g_k^*(t)\\
      \end{pmatrix}
      = \begin{pmatrix}
         -\beta_k & -\beta_k\\ 
         \alpha_k + \lambda_k & \alpha_k-\lambda_k\\ 
      \end{pmatrix}
      \begin{pmatrix}
         i\lambda_k & 0\\ 
         0 & -i\lambda_k\\ 
      \end{pmatrix}
      \begin{pmatrix}
         \frac{\alpha_k-\lambda_k}{2\beta_k \lambda_k} & \frac{1}{2 \lambda_k}\\ 
         \frac{-\alpha_k-\lambda_k}{2\beta_k \lambda_k} & \frac{-1}{2 \lambda_k}\\ 
      \end{pmatrix}
      \begin{pmatrix}
         f_k(t)\\ 
         g_k^*(t)\\
      \end{pmatrix},
   \end{aligned}
\end{equation}
where
\begin{equation}
   \begin{aligned}
      \begin{pmatrix}
         -\beta_k & -\beta_k\\ 
         \alpha_k +\lambda_k & \alpha_k -\lambda_k\\ 
      \end{pmatrix}
       = 
       \begin{pmatrix}
         \frac{\alpha_k -\lambda_k}{2\beta_k  \lambda_k} & \frac{1}{2 \lambda_k}\\ 
         \frac{-\alpha_k -\lambda_k}{2\beta_k \lambda_k} & \frac{-1}{2 \lambda_k}\\ 
      \end{pmatrix}^{-1}.
   \end{aligned}
\end{equation}
The differential equation \eqref{Eq.: diagonal diff.eq.} can then be solved as
\begin{equation}\label{Eq.: fg Matrix}
   \begin{aligned}
      \begin{pmatrix}
         f_k(t)\\ 
         g_k^*(t)\\
      \end{pmatrix}
      &=
      \begin{pmatrix}
         -\beta_k & -\beta_k\\ 
         \alpha_k +\lambda_k & \alpha_k-\lambda_k\\ 
      \end{pmatrix}
      \begin{pmatrix}
         e^{i\lambda_k (t-t_0)} & 0\\ 
         0 & e^{-i\lambda_k (t-t_0)}\\ 
      \end{pmatrix}
      \begin{pmatrix}
         \frac{\alpha_k-\lambda_k}{2\beta_k \lambda_k} & \frac{1}{2 \lambda_k}\\ 
         \frac{-\alpha_k-\lambda_k}{2\beta_k \lambda_k} & \frac{-1}{2 \lambda_k}\\ 
      \end{pmatrix}
      \begin{pmatrix}
         f_k(t_0)\\ 
         g_k^*(t_0)\\
      \end{pmatrix}\\
      &=
      \begin{pmatrix}
         \cos(\lambda_k (t-t_0)) - \frac{\alpha_k i}{\lambda_k}\sin(\lambda_k (t-t_0)) & -\frac{\beta_k i}{\lambda_k} \sin(\lambda_k (t-t_0))\\ 
         \frac{\beta_k i}{\lambda_k} \sin(\lambda_k (t-t_0)) & \cos(\lambda_k (t-t_0)) + \frac{\alpha_k i}{\lambda_k}\sin(\lambda_k (t-t_0))\\
      \end{pmatrix}
      \begin{pmatrix}
         f_k(t_0)\\ 
         g_k^*(t_0)\\
      \end{pmatrix}\\
      &=
      \begin{pmatrix}
         \cos(\frac{\omega_k(t-t_0)}{n_\rm{r}}) - \frac{i}{2}\left(\eta_\rm{r} + \eta_\rm{r}^{-1}\right)\sin(\frac{\omega_k(t-t_0)}{n_\rm{r}}) & -\frac{(-1)^\lambda i}{2}\left(\eta_\rm{r} - \eta_\rm{r}^{-1}\right) \sin(\frac{\omega_k(t-t_0)}{n_\rm{r}})\\ 
         \frac{(-1)^\lambda i}{2}\left(\eta_\rm{r} - \eta_\rm{r}^{-1}\right) \sin(\frac{\omega_k(t-t_0)}{n_\rm{r}}) & \cos(\frac{\omega_k(t-t_0)}{n_\rm{r}}) + \frac{i}{2}\left(\eta_\rm{r} + \eta_\rm{r}^{-1}\right) \sin(\frac{\omega_k(t-t_0)}{n_\rm{r}})\\
      \end{pmatrix}
      \begin{pmatrix}
         f_k(t_0)\\ 
         g_k^*(t_0)\\
      \end{pmatrix},
   \end{aligned}
\end{equation}
with $\eta_\rm{r} = \sqrt{\frac{\mu_\rm{r}}{\epsilon_\rm{r}}}$ and $n_\rm{r} = \sqrt{\epsilon_\rm{r} \mu_\rm{r}}$.
If we now choose $t_0 = 0$ and use the initial conditions $f_k(0) = 1$ and $g_k(0)=0$, we find the analytical solution to a temporal step to be
\begin{equation}
   \left\{
       \begin{aligned}
           f_k(t) &= \cos(\frac{\omega_k t}{n_\rm{r}}) - \frac{i}{2}\left(\eta_\rm{r} + \eta_\rm{r}^{-1}\right)\sin(\frac{\omega_k t}{n_\rm{r}}),\\ 
         g_k(t) &= -\frac{(-1)^\lambda i}{2}\left(\eta_\rm{r} - \eta_\rm{r}^{-1}\right) \sin(\frac{\omega_k t}{n_\rm{r}}).
       \end{aligned}
   \right.
\end{equation}
Notice that for an impedance-matched modulation, so $\sqrt{\frac{\mu_2}{\epsilon_2}} = \sqrt{\frac{\mu_1}{\epsilon_1}}$, we find $\mu_r = \epsilon_r$, which implies $g_k(t) = 0$ and $f_k(t) = e^{-i\frac{\omega_k }{\sqrt{\epsilon_\rm{r} \mu_\rm{r}}}t}$.
In this case, the time-dependent annihilation operator becomes $\hat{a}_\b{k}(t) = e^{-i\frac{\omega_k }{\sqrt{\epsilon_\rm{r} \mu_\rm{r}}}t} \hat{a}_\b{k}(0)$ which implies that the number operator $\hat{n}_\b{k}(t) = \hat{a}^\dagger_\b{k}(t) \hat{a}_\b{k}(t) = \hat{a}^\dagger_\b{k}(0) \hat{a}_\b{k}(0)$ is conserved over time, indicating no photons are created or annihilated.
This corresponds to the classical result that an impedance-matched modulation does not lead to any backscattering of classical light~\cite{morgenthaler_velocity_1958}.

\newpage
\section{Initial Vacuum State Evolution [Eq.~(15)]}
When there are initially no photons present, $\ket{\psi(0)} = \ket{0,0}$, the state of the system at later times is described by $\ket{\psi(t)} = \sum_{n, m} C_{n, m}(t) \ket{n, m}$.
As explained in the paper, the coefficients $C_{n, m}(t)$ can be calculated as
\begin{equation}
   \begin{aligned}
      C_{n, m}(t) &= \frac{1}{\sqrt{n!m!}} \left\langle \xi_{0,0}(t) \left| \hat{a}^n_{\b{k}, \lambda}(t) \hat{a}^m_{-\b{k}, \lambda}(t) \right| 0, 0 \right\rangle
   \end{aligned}
\end{equation}
With the time-dependent annihilation operators $\hat{a}_{\pm \b{k}, \lambda}(t) = f_k(t)\hat{a}_{\pm \b{k}, \lambda} + g_k(t)\hat{a}^\dagger_{\mp \b{k}, \lambda}$ and the binomial formula, 
we find
\begin{equation}
   \begin{aligned}
      C_{n, m}(t) &= \frac{1}{\sqrt{n!m!}} \sum_{i=0}^n \sum_{j=0}^m \binom{n}{i} \binom{m}{j} f_k^{n-i+m-j} g_k^{i+j} 
      \bra{\xi_{0,0}(t)} \hat{a}^{n-i}_{\b{k}, \lambda} (\hat{a}^\dagger_{-\b{k}, \lambda})^i \hat{a}^{m-j}_{-\b{k}, \lambda} (\hat{a}^\dagger_{\b{k}, \lambda})^j \ket{0, 0}.
   \end{aligned}
\end{equation}
Since $\hat{a}_{-\b{k}, \lambda} \ket{0, 0} = 0$, the sum over $j$ retains only the term with $j=m$ and 
\begin{equation}\label{Eq.: C_nm -1}
   \begin{aligned}
      C_{n, m}(t) &= \frac{1}{\sqrt{n!m!}} \sum_{i=0}^n \binom{n}{i} f_k^{n-i} g_k^{m+i} T,
   \end{aligned}
\end{equation}
with $T = \bra{\xi_{0,0}(t)} \hat{a}^{n-i}_{\b{k}, \lambda} (\hat{a}^\dagger_{-\b{k}, \lambda})^i (\hat{a}^\dagger_{\b{k}, \lambda})^m \ket{0, 0}$.
If we apply the creation operators on the vacuum state in $T$, we find $T = \sqrt{m! i!} \bra{\xi_{0,0}(t)} \hat{a}^{n-i}_{\b{k}, \lambda} \ket{m, i}$.
Now we apply the annihilation operators on $\ket{m, i}$ so that
\begin{equation}
   \begin{aligned}
      T = \sqrt{m! i!} \sqrt{\frac{m!}{(m-(n-i)!)}} \bra{\xi_{0,0}(t)} \ket{m-(n-i), i} \Theta(m-(n-i)).
   \end{aligned}
\end{equation}
The Heaviside step function $\Theta(m-(n-i))$ appears because for $m<n-i$ we apply more annihilation operators than there are photons in the forward mode, which results in zero.
We now use \eqref{Eq.: Instantaneous Ground State} and the orthonormality of the Fock states to find
\begin{equation}
   \begin{aligned}
      T = m! \left(-\frac{g_k^i(t)}{f_k^{i+1}(t)}\right)^* e^{-i\theta(t)} \delta_{n, m}.
   \end{aligned}
\end{equation}
We insert $T$ back into \eqref{Eq.: C_nm -1} and use the binomial formula and $|f_k(t)|^2 - |g_k(t)|^2 = 1$, to find
\begin{equation}\label{Eq.: C_nm phase}
   \begin{aligned}
      C_{n, m}(t) = \frac{g_k^n(t)}{(f^{n+1}_k(t))^*} e^{-i\theta(t)} \delta_{n, m}.
   \end{aligned}
\end{equation}
To determine the phase factor $\theta(t)$, we go back to the Schrödinger equation.
According to \eqref{Eq.: Schrödinger Equation}, the differential equations for $C_{n, n}(t)$ are
\begin{equation}\label{Eq.: Schrodinger Cnn}
   \begin{aligned}
    i \partial_t C_{n,n}(t) &= \alpha_k(t) \left( 2n + 1 \right) C_{n, n}(t) + \beta_k(t) (n+1) C_{n+1, n+1}(t) + \beta_k(t) n C_{n-1, n-1}(t).
   \end{aligned}
\end{equation}
The time derivative of $C_{n, n}(t)$ is
\begin{equation}\label{Eq.: partial Cnn -1}
   \begin{aligned}
    \partial_t C_{n,n}(t) &= \left(\frac{n g^{n-1}_k(t)}{(f_k^{n+1}(t))^*} \partial_t g_k(t) - (n+1)\frac{g^n_k(t)}{(f_k^{n+2}(t))^*}\partial_t f^*_k(t)
    -i \frac{g^n_k(t)}{(f_k^{n+1}(t))^*} \partial_t \theta(t)\right) e^{-i\theta(t)}.
   \end{aligned}
\end{equation}
The functions $f_k$ and $g_k$ are defined by the differential equations
\begin{equation}
   \left\{
       \begin{aligned}
        \partial_t f_k(t) &= -i\alpha_k(t)f_k(t) -i  \beta_k(t)g_k^*(t), \quad f_k(0) = 1,\\
        \partial_t g_k(t) &= -i\alpha_k(t)g_k(t) -i \beta_k(t)f^*_k(t), \quad g_k(0) = 0.\\
       \end{aligned}
   \right.
\end{equation}
If we insert these equations into \eqref{Eq.: partial Cnn -1}, and then use \eqref{Eq.: C_nm phase}, we find
\begin{equation}\label{Eq.: partial Cnn}
   \begin{aligned}
   \hspace{-2.7mm} i \partial_t C_{n,n}(t) \hspace{-0.7mm} &= \hspace{-0.7mm} [ \alpha_k(t) \hspace{-0.71mm} \left( 2n \hspace{-0.4mm} + \hspace{-0.4mm} 1 \right) \hspace{-0.7mm} C_{n, n}(t) \hspace{-0.7mm} + \hspace{-0.7mm} \beta_k(t) (n \hspace{-0.4mm} + \hspace{-0.4mm} 1) C_{n+1, n+1}(t) 
    \hspace{-0.7mm} + \hspace{-0.7mm} \beta_k(t) n C_{n-1, n-1}(t)
    \hspace{-0.7mm} - \hspace{-0.7mm} i C_{n,n}(t) \partial_t \theta(t) ] e^{ \hspace{-1mm} -i\theta(t)} \hspace{-0.7mm} .
   \end{aligned}
\end{equation}
We can then compare \eqref{Eq.: Schrodinger Cnn} and \eqref{Eq.: partial Cnn}, we find that the phase factor $\theta(t) = 0$.
Therefore, the instantaneous ground state in \eqref{Eq.: Instantaneous Ground State} is now completely determined and the vacuum state evolves as 
\begin{equation}
   \begin{aligned}
      \ket{\psi(t)} = \sum_{n} \frac{g^n_k(t)}{(f_k^{n+1}(t))^*} \ket{n, n}.
   \end{aligned}
\end{equation}

\section{Designing Modulation profiles for an ATC}
The goal of an antireflection temporal coating (ATC) is to change the refractive index of a medium in time so that the frequency of an incoming signal, $\omega_1$, is converted to $\omega_2 = \frac{n_2}{n_1}\omega_1$, without generating additional photons or noise.
We will first explain how this noise can be calculated, and then present a method to determine a suitable temporal modulation profile of the medium to minimize this noise.
First of all, the Hamiltonian in \eqref{Eq.: H(a, ad, t)S} shows that if we change the permittivity and permeability of a medium from $\epsilon_1$ and $\mu_1$ to $\epsilon_2$ and $\mu_2$, the Hamiltonian becomes 
\begin{equation}\label{Eq.: H SS}
   \begin{aligned}
   \hat{H} = \int_V \rm{d}^3 r \mathcal{H}(t) &= \sum_{\bf{k}, \lambda} \hbar \alpha_k \left(   \hat{n}_{\bf{k}, \lambda} + \frac{1}{2}     \right)
   + \frac{\hbar \beta_k}{2} \left(\hat{a}_{\bf{k}, \lambda}  \hat{a}_{-\bf{k}, \lambda}
   +
   \hat{a}^\dagger_{\bf{k}, \lambda}(t) \hat{a}^\dagger_{-\bf{k}, \lambda} \right),\\
   \end{aligned}
\end{equation}
with $\alpha_k = \frac{\omega_k}{2}\left(\frac{\epsilon_1}{\epsilon_2} + \frac{\mu_1}{\mu_2}\right)$ and $\beta_k(t) = \frac{(-1)^\lambda \omega_k}{2}\left(\frac{\epsilon_1}{\epsilon_2} - \frac{\mu_1}{\mu_2}\right)$.
For a modulation that is not impedance-matched, this means that $\beta_k \neq 0$ so pairwise creation and annihilaton events remain present.
To eliminate these interaction terms, we perform a Bogoliubov transformation
\begin{equation}\label{Eq.: Bogoliubov I}
   \left\{
       \begin{aligned}
           \hat{a}_{\bf{k}, \lambda} &= c \hat{b}_{\bf{k}, \lambda} + d \hat{b}^{\dagger}_{-\bf{k}, \lambda},\\
            \hat{a}^\dagger_{-\bf{k}, \lambda} &= c \hat{b}^\dagger_{-\bf{k}, \lambda} + d \hat{b}_{\bf{k}, \lambda},\\
       \end{aligned}
   \right.
\end{equation}
where $c = \frac{1}{2}\left(\sqrt{\frac{\eta_1}{\eta_2}}+\sqrt{\frac{\eta_2}{\eta_1}}\right)$, $d = \frac{(-1)^\lambda}{2}\left(\sqrt{\frac{\eta_1}{\eta_2}}-\sqrt{\frac{\eta_2}{\eta_1}}\right)$, the impedance is $\eta_i = \sqrt{\mu_i/\epsilon_i}$ and the operators $\hat{b}_{\bf{k}, \lambda}$ satisfy $[\hat{b}_{\bf{k}, \lambda}, \hat{b}^\dagger_{\bf{q}, \nu}] = \delta_{\bf{k}, \bf{q}}\delta_{\lambda, \nu}$.
Inserting these Bogoliubov transformations into the Hamiltonian in \eqref{Eq.: H SS}, we find
\begin{equation}\label{Eq.: H Final}
   \begin{aligned}
      \hat{H} &= \sum_{\bf{k}, \lambda} \hbar \Omega_k \left(\hat{b}^\dagger_{\bf{k}, \lambda} \hat{b}_{\bf{k}, \lambda} + \frac{1}{2} \right),
   \end{aligned}
\end{equation}
where $\Omega_k = |\bf{k}|/\sqrt{\epsilon_2 \mu_2}$.
Fock states generated by the operators $\hat{b}^\dagger_{\bf{k}, \lambda}$ are then eigenstates of this Hamiltonian, which makes them more relevant as these states only change by a phase factor as time goes on.
If we inverse the relations in \eqref{Eq.: Bogoliubov I}, we find 
\begin{equation}\label{Eq.: Bogoliubov II}
   \left\{
       \begin{aligned}
           \hat{b}_{\bf{k}, \lambda} &= c \hat{a}_{\bf{k}, \lambda} + d \hat{a}^{\dagger}_{-\bf{k}, \lambda},\\
            \hat{b}^\dagger_{-\bf{k}, \lambda} &= c \hat{a}^\dagger_{-\bf{k}, \lambda} + d \hat{a}_{\bf{k}, \lambda}.\\
       \end{aligned}
   \right.
\end{equation}
In the Heisenberg picture, we then have 
\begin{equation}
   \left\{
       \begin{aligned}
           \hat{b}_{\bf{k}, \lambda}(t) &= c \hat{a}_{\bf{k}, \lambda}(t) + d \hat{a}^{\dagger}_{-\bf{k}, \lambda}(t),\\
            \hat{b}^\dagger_{-\bf{k}, \lambda}(t) &= c \hat{a}^\dagger_{-\bf{k}, \lambda}(t) + d \hat{a}_{\bf{k}, \lambda}(t),\\
       \end{aligned}
   \right.
\end{equation}
and if we insert $\hat{a}_{\pm \b{k}, \lambda}(t) = f_k(t)\hat{a}_{\pm \b{k}, \lambda} + g_k(t)\hat{a}^\dagger_{\mp \b{k}, \lambda}$, we find 
\begin{equation}\label{Eq.: Bogoliubov III}
   \left\{
       \begin{aligned}
           \hat{b}_{\bf{k}, \lambda}(t) &= F_k(t) \hat{a}_{\bf{k}, \lambda}(0) + G_k(t) \hat{a}^{\dagger}_{-\bf{k}, \lambda}(0),\\
            \hat{b}^\dagger_{-\bf{k}, \lambda}(t) &= F_k^*(t) \hat{a}^\dagger_{-\bf{k}, \lambda}(0) + G_k^*(t) \hat{a}_{\bf{k}, \lambda}(0),\\
       \end{aligned}
   \right.
\end{equation}
where $F_k(t) = cf_k(t) - dg_k^*(t)$ and $G_k(t) = -d f_k^*(t) + c g_k(t)$.
We can then express the number operator in the second medium as 
\begin{equation}
   \begin{aligned}
      \hat{b}^\dagger_{\bf{k}, \lambda}(t)\hat{b}_{\bf{k}, \lambda}(t) = |F_k(t)|^2 \hat{n}_{\bf{k},\lambda}(0) + |G_k(t)|^2 (\hat{n}_{-\bf{k},\lambda}(0)+1) + F_k(t) G_k(t) \hat{a}^\dagger_{\bf{k}, \lambda}(0) \hat{a}^{\dagger}_{-\bf{k}, \lambda}(0) + F^*_k(t) G^*_k(t) \hat{a}_{\bf{k}, \lambda}(0) \hat{a}_{-\bf{k}, \lambda}(0).
   \end{aligned}
\end{equation}
Now, assuming we start with a single photon $\ket{1_{\bf{k}, \lambda}}$, we find that in the final medium the number of photons with that momentum and polarization is $\bra{1_{\bf{k}, \lambda}} \hat{b}^\dagger_{\bf{k}, \lambda}(t)\hat{b}_{\bf{k}, \lambda}(t) \ket{1_{\bf{k}, \lambda}} = |F_k(t)|^2 + |G_k(t)|^2$, and because \eqref{Eq.: Bogoliubov III} is also a Bogoliubov transformation, $|F_k(t)|^2-|G_k(t)|^2 = 1$, and the photon number in the final medium will be 
\begin{equation}\label{Eq.: Produced Noise}
   \begin{aligned}
      \bra{1_{\bf{k}, \lambda}} \hat{b}^\dagger_{\bf{k}, \lambda}(t)\hat{b}_{\bf{k}, \lambda}(t) \ket{1_{\bf{k}, \lambda}} = 1 + 2|G_k(t)|^2.
   \end{aligned}
\end{equation}
Notice that because the number operator in the final medium $\hat{b}^\dagger_{\bf{k}, \lambda}(t)\hat{b}_{\bf{k}, \lambda}(t)$ commutes with the Hamiltonian in that medium, \eqref{Eq.: H Final}, the photon number remains fixed after the modulation has ended.
We then find that $|G_k(t)|^2$ will quantify the produced noise of an ATC when converting photons of frequency $\omega_k = |\bf{k}|/\sqrt{\epsilon_1 \mu_1}$ to $\Omega_k = |\bf{k}|/\sqrt{\epsilon_2 \mu_2}$.
An ideal ATC would then result in $G_k(t) = 0$ at the end of the procedure.
Now if we want an ATC that operates without noise for a band of frequencies $[\omega_\rm{min}, \omega_\rm{max}]$, it is useful to define the ATC quality factor as
\begin{equation}
   \begin{aligned}
      Q = \frac{1}{\omega_\rm{max}-\omega_\rm{min}}\int_{\omega_\rm{min}}^{\omega_\rm{max}} d\omega_k |G_k(t_f)|^2,
   \end{aligned}
\end{equation}
where $t_f$ is the time at which the medium modulation ends.
An ideal ATC in the range of $[\omega_\rm{min}, \omega_\rm{max}]$ has $Q=0$, while higher values of $Q$ indicate more noisy ATCs.

\begin{figure}[!h]
    \centering
    \includegraphics[width=0.75\linewidth]{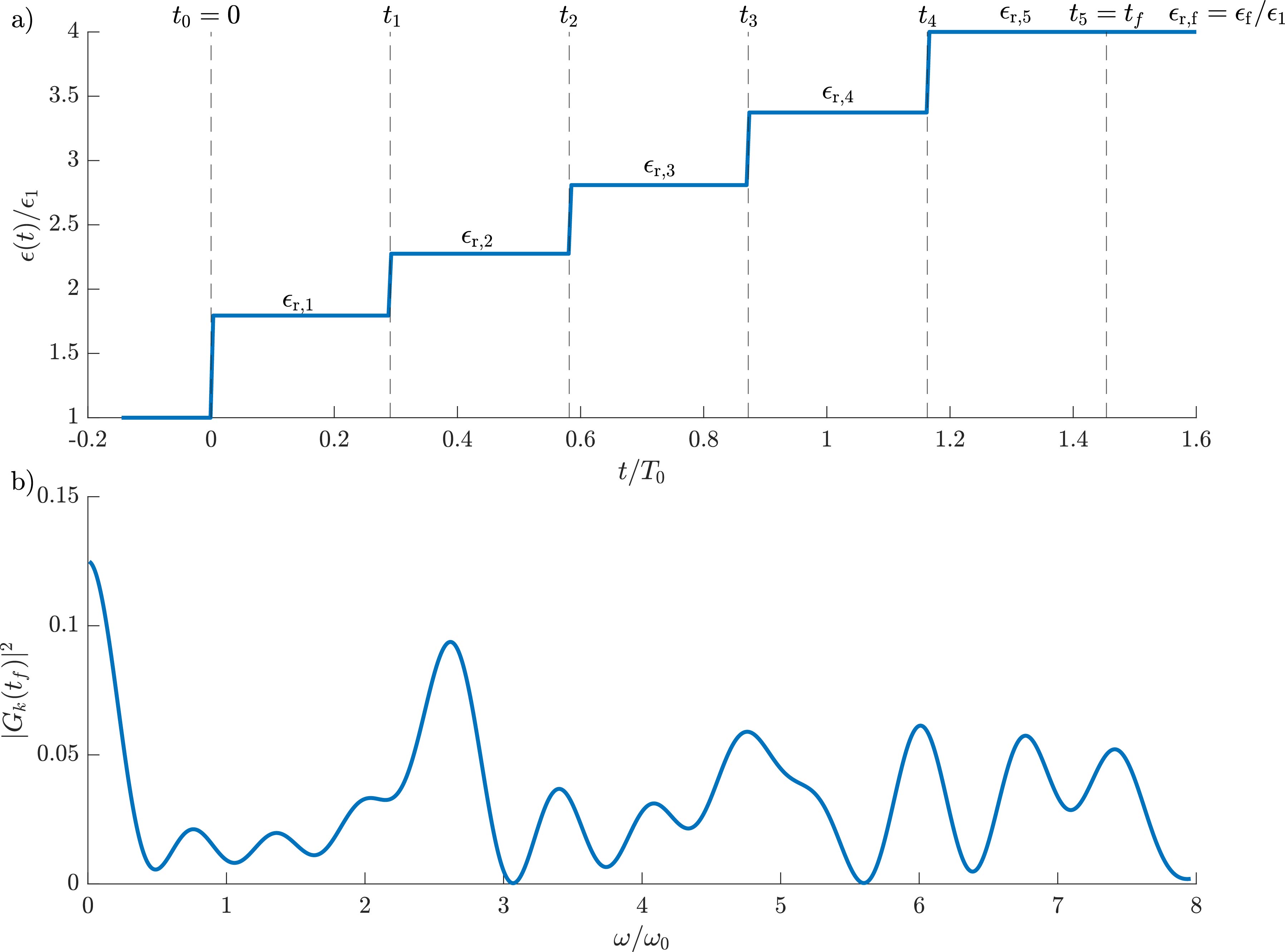}
    \caption{Illustration of the method used to determine the permittivity profile for an optimal ATC, shown here for constant permeability and piecewise-constant permittivity. The total operation time $t_f$ is divided into $n$ equal intervals of duration $dt = t_f/n$, with the permittivity assumed constant within each interval but allowed to vary between intervals.
    b) Corresponding $|G_k(t_f)|^2$ as a function of the initial signal frequency $\omega$ for the permittivity modulation shown in a).}
    \label{Fig.: MethodDemonstration}
\end{figure}

\begin{figure}[!h]
    \centering
    \includegraphics[width=\linewidth]{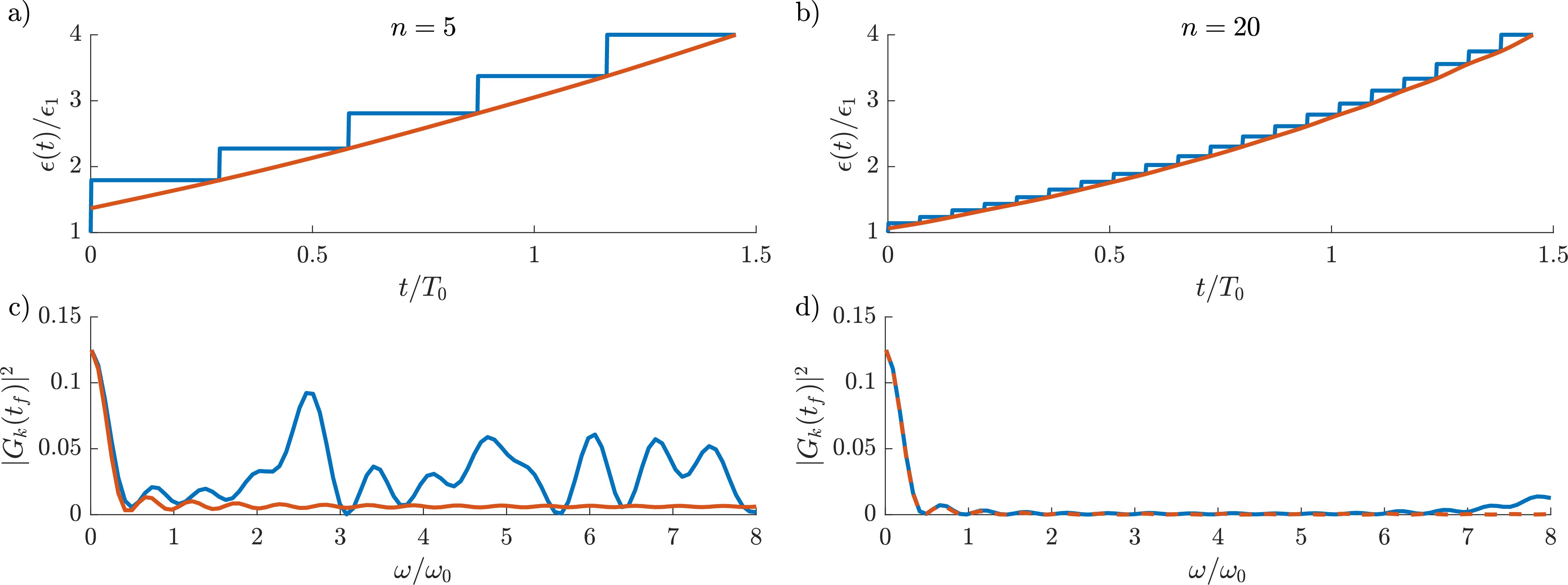}
    \caption{a, b) Stepwise permittivity modulations minimizing $Q$ for $n=5$ and $n=20$ (blue) and their spline interpolations (orange).
    c, d) Corresponding $|G_k(t_f)|^2(\omega)$ for the stepwise (blue) and spline interpolated (orange) modulations.}
    \label{Fig.: SplineFit}
\end{figure}

Since $G_k(t_f)$ depends on $f_k(t_f)$ and $g_k(t_f)$, which need to be determined by the differential equations in \eqref{Eq.: diagonal diff.eq.}, for which solutions are difficult to find analytically for general temporal modulations, it is difficult to assess which types of modulations would lead to high-quality ATCs. 
The only exception for this problem are modulations consisting out of a series of temporal steps, for $f_k(t_f)$ and $g_k(t_f)$ can be calculated with \eqref{Eq.: fg Matrix}. 
Therefore, we propose the following method for calculating the required temporal modulations for high-quality ATCs.
First, we assume the signal arrives at the ATC at time $t=0$ and should be converted at $t=t_f$, and we divide this time interval into $n$ parts of time $dt = t_f/n$.
Second, we assume that in each of these intervals, the permittivity and permeability are constant, but their values can be different in different intervals, this is illustrated in \figref{Fig.: MethodDemonstration}.
If we define $t_m = (m/n) t_f$ for $m\in{0, 1, ..., n}$ and $n_{\rm{r}, m}, \eta_{\rm{r}, m}$ the refractive index and impedance in the interval $[t_{m-1}, t_m]$, \eqref{Eq.: fg Matrix} tells us that 
\begin{equation}
   \begin{aligned}
      \begin{pmatrix}
         f_k(t_m)\\ 
         g_k^*(t_m)\\
      \end{pmatrix}
      &=
      \begin{pmatrix}
         \cos(\frac{\omega_k dt}{n_{\rm{r}, m}}) - \frac{i}{2}\left(\eta_{\rm{r}, m} + \eta_{\rm{r}, m}^{-1}\right)\sin(\frac{\omega_k dt}{n_{\rm{r}, m}}) & -\frac{(-1)^\lambda i}{2}\left(\eta_{\rm{r}, m} - \eta_{\rm{r}, m}^{-1}\right) \sin(\frac{\omega_k dt}{n_{\rm{r}, m}})\\ 
         \frac{(-1)^\lambda i}{2}\left(\eta_{\rm{r}, m} - \eta_{\rm{r}, m}^{-1}\right) \sin(\frac{\omega_k dt}{n_{\rm{r}, m}}) & \cos(\frac{\omega_k dt}{n_{\rm{r}, m}}) + \frac{i}{2}\left(\eta_{\rm{r}, m} + \eta_{\rm{r}, m}^{-1}\right) \sin(\frac{\omega_k dt}{n_{\rm{r}, m}})\\
      \end{pmatrix}
      \begin{pmatrix}
         f_k(t_{m-1})\\ 
         g_k^*(t_{m-1})\\
      \end{pmatrix}.
   \end{aligned}
\end{equation}
As $t_f = t_n$, we can use this equation recursively to determine $f_k(t_f)$ and $g^*_k(t_f)$ in function of the different permittivity and permeability steps encoded in $n_{\rm{r}, m}$ and $\eta_{\rm{r}, m}$, and $f_k(t_0)=1$ and $g_k^*(t_0)=0$ as $t_0 = 0$.
With these values of $f_k(t_f)$ and $g^*_k(t_f)$, we can then calculate $G_k(t_f)$, which allows us to determine the quality factor of the ATC as a function of the permittivity and permeability steps encoded in $n_{\rm{r}, m}$ and $\eta_{\rm{r}, m}$.
Then, we use built-in methods in Matlab to minimize $Q$ as a function of the permittivity and permeability steps, as we increase the size of $n$, we would then expect the modulation profiles to look increasingly continuous.
Finally, we will use spline interpolation to find continuous modulation profiles resembling the stepwise modulation profiles.
This method is illustrated in \figref{Fig.: SplineFit} for an ATC designed with a frequency range of $[0.01\omega_0, 6\omega_0]$, an operation time of $1.454T_0$ with $T_0 = 2\pi/\omega_0$. For simplicity, this example assumes the permeability to be constant, and only the permittivity is varied in time.
The figure in the main paper, Fig. (3), corresponds to the plots b) and d) in \figref{Fig.: SplineFit} found below, where the amount of produced photons equals $2|G_k(t_f)|$ as presented in \eqref{Eq.: Produced Noise} 
.
\newpage

%% file: main.bbl
\begin{thebibliography}{50}%
\makeatletter
\providecommand \@ifxundefined [1]{%
 \@ifx{#1\undefined}
}%
\providecommand \@ifnum [1]{%
 \ifnum #1\expandafter \@firstoftwo
 \else \expandafter \@secondoftwo
 \fi
}%
\providecommand \@ifx [1]{%
 \ifx #1\expandafter \@firstoftwo
 \else \expandafter \@secondoftwo
 \fi
}%
\providecommand \natexlab [1]{#1}%
\providecommand \enquote  [1]{``#1''}%
\providecommand \bibnamefont  [1]{#1}%
\providecommand \bibfnamefont [1]{#1}%
\providecommand \citenamefont [1]{#1}%
\providecommand \href@noop [0]{\@secondoftwo}%
\providecommand \href [0]{\begingroup \@sanitize@url \@href}%
\providecommand \@href[1]{\@@startlink{#1}\@@href}%
\providecommand \@@href[1]{\endgroup#1\@@endlink}%
\providecommand \@sanitize@url [0]{\catcode `\\12\catcode `\$12\catcode
  `\&12\catcode `\#12\catcode `\^12\catcode `\_12\catcode `\%12\relax}%
\providecommand \@@startlink[1]{}%
\providecommand \@@endlink[0]{}%
\providecommand \url  [0]{\begingroup\@sanitize@url \@url }%
\providecommand \@url [1]{\endgroup\@href {#1}{\urlprefix }}%
\providecommand \urlprefix  [0]{URL }%
\providecommand \Eprint [0]{\href }%
\providecommand \doibase [0]{https://doi.org/}%
\providecommand \selectlanguage [0]{\@gobble}%
\providecommand \bibinfo  [0]{\@secondoftwo}%
\providecommand \bibfield  [0]{\@secondoftwo}%
\providecommand \translation [1]{[#1]}%
\providecommand \BibitemOpen [0]{}%
\providecommand \bibitemStop [0]{}%
\providecommand \bibitemNoStop [0]{.\EOS\space}%
\providecommand \EOS [0]{\spacefactor3000\relax}%
\providecommand \BibitemShut  [1]{\csname bibitem#1\endcsname}%
\let\auto@bib@innerbib\@empty
\bibitem [{\citenamefont {Louisell}\ \emph {et~al.}(1961)\citenamefont
  {Louisell}, \citenamefont {Yariv},\ and\ \citenamefont
  {Siegman}}]{louisell_quantum_1961}%
  \BibitemOpen
  \bibfield  {author} {\bibinfo {author} {\bibfnamefont {W.~H.}\ \bibnamefont
  {Louisell}}, \bibinfo {author} {\bibfnamefont {A.}~\bibnamefont {Yariv}},\
  and\ \bibinfo {author} {\bibfnamefont {A.~E.}\ \bibnamefont {Siegman}},\
  }\bibfield  {title} {\bibinfo {title} {Quantum {Fluctuations} and {Noise} in
  {Parametric} {Processes}. {I}.},\ }\href@noop {} {\bibfield  {journal}
  {\bibinfo  {journal} {Physical Review}\ }\textbf {\bibinfo {volume} {124}},\
  \bibinfo {pages} {1646} (\bibinfo {year} {1961})}\BibitemShut {NoStop}%
\bibitem [{\citenamefont {Gordon}\ \emph {et~al.}(1963)\citenamefont {Gordon},
  \citenamefont {Louisell},\ and\ \citenamefont
  {Walker}}]{gordon_quantum_1963}%
  \BibitemOpen
  \bibfield  {author} {\bibinfo {author} {\bibfnamefont {J.~P.}\ \bibnamefont
  {Gordon}}, \bibinfo {author} {\bibfnamefont {W.~H.}\ \bibnamefont
  {Louisell}},\ and\ \bibinfo {author} {\bibfnamefont {L.~R.}\ \bibnamefont
  {Walker}},\ }\bibfield  {title} {\bibinfo {title} {Quantum {Fluctuations} and
  {Noise} in {Parametric} {Processes}. {II}},\ }\href@noop {} {\bibfield
  {journal} {\bibinfo  {journal} {Physical Review}\ }\textbf {\bibinfo {volume}
  {129}},\ \bibinfo {pages} {481} (\bibinfo {year} {1963})}\BibitemShut
  {NoStop}%
\bibitem [{\citenamefont {Mendonça}\ \emph {et~al.}(2000)\citenamefont
  {Mendonça}, \citenamefont {Guerreiro},\ and\ \citenamefont
  {Martins}}]{mendonca_quantum_2000}%
  \BibitemOpen
  \bibfield  {author} {\bibinfo {author} {\bibfnamefont {J.~T.}\ \bibnamefont
  {Mendonça}}, \bibinfo {author} {\bibfnamefont {A.}~\bibnamefont
  {Guerreiro}},\ and\ \bibinfo {author} {\bibfnamefont {A.~M.}\ \bibnamefont
  {Martins}},\ }\bibfield  {title} {\bibinfo {title} {Quantum theory of time
  refraction},\ }\href {https://doi.org/10.1103/PhysRevA.62.033805} {\bibfield
  {journal} {\bibinfo  {journal} {Physical Review A}\ }\textbf {\bibinfo
  {volume} {62}},\ \bibinfo {pages} {033805} (\bibinfo {year}
  {2000})}\BibitemShut {NoStop}%
\bibitem [{\citenamefont {Horsley}\ and\ \citenamefont
  {Pendry}(2023)}]{Horsley_Pendry_2023}%
  \BibitemOpen
  \bibfield  {author} {\bibinfo {author} {\bibfnamefont {S.~A.~R.}\
  \bibnamefont {Horsley}}\ and\ \bibinfo {author} {\bibfnamefont {J.~B.}\
  \bibnamefont {Pendry}},\ }\bibfield  {title} {\bibinfo {title} {Quantum
  electrodynamics of time-varying gratings},\ }\href
  {https://doi.org/10.1073/pnas.2302652120} {\bibfield  {journal} {\bibinfo
  {journal} {Proceedings of the National Academy of Sciences}\ }\textbf
  {\bibinfo {volume} {120}},\ \bibinfo {pages} {e2302652120} (\bibinfo {year}
  {2023})}\BibitemShut {NoStop}%
\bibitem [{\citenamefont {Moore}(1970)}]{moore_quantum_1970}%
  \BibitemOpen
  \bibfield  {author} {\bibinfo {author} {\bibfnamefont {G.~T.}\ \bibnamefont
  {Moore}},\ }\bibfield  {title} {\bibinfo {title} {Quantum {Theory} of the
  {Electromagnetic} {Field} in a {Variable}-{Length} {One}-{Dimensional}
  {Cavity}},\ }\href {https://doi.org/10.1063/1.1665432} {\bibfield  {journal}
  {\bibinfo  {journal} {Journal of Mathematical Physics}\ }\textbf {\bibinfo
  {volume} {11}},\ \bibinfo {pages} {2679} (\bibinfo {year}
  {1970})}\BibitemShut {NoStop}%
\bibitem [{\citenamefont {Hui}\ \emph {et~al.}(2000)\citenamefont {Hui},
  \citenamefont {Qing-Yun},\ and\ \citenamefont {Jian-Sheng}}]{HUI2000174}%
  \BibitemOpen
  \bibfield  {author} {\bibinfo {author} {\bibfnamefont {J.}~\bibnamefont
  {Hui}}, \bibinfo {author} {\bibfnamefont {S.}~\bibnamefont {Qing-Yun}},\ and\
  \bibinfo {author} {\bibfnamefont {W.}~\bibnamefont {Jian-Sheng}},\ }\bibfield
   {title} {\bibinfo {title} {Dynamical casimir effect at finite temperature},\
  }\href {https://doi.org/https://doi.org/10.1016/S0375-9601(00)00165-1}
  {\bibfield  {journal} {\bibinfo  {journal} {Physics Letters A}\ }\textbf
  {\bibinfo {volume} {268}},\ \bibinfo {pages} {174} (\bibinfo {year}
  {2000})}\BibitemShut {NoStop}%
\bibitem [{\citenamefont {Dodonov}(2009)}]{dodonov_dynamical_2009}%
  \BibitemOpen
  \bibfield  {author} {\bibinfo {author} {\bibfnamefont {V.~V.}\ \bibnamefont
  {Dodonov}},\ }\bibfield  {title} {\bibinfo {title} {Dynamical {Casimir}
  effect: {Some} theoretical aspects},\ }\href
  {https://doi.org/10.1088/1742-6596/161/1/012027} {\bibfield  {journal}
  {\bibinfo  {journal} {Journal of Physics: Conference Series}\ }\textbf
  {\bibinfo {volume} {161}},\ \bibinfo {pages} {012027} (\bibinfo {year}
  {2009})}\BibitemShut {NoStop}%
\bibitem [{\citenamefont {Román-Ancheyta}\ \emph {et~al.}(2017)\citenamefont
  {Román-Ancheyta}, \citenamefont {Ramos-Prieto}, \citenamefont {Perez-Leija},
  \citenamefont {Busch},\ and\ \citenamefont
  {León-Montiel}}]{roman-ancheyta_dynamical_2017}%
  \BibitemOpen
  \bibfield  {author} {\bibinfo {author} {\bibfnamefont {R.}~\bibnamefont
  {Román-Ancheyta}}, \bibinfo {author} {\bibfnamefont {I.}~\bibnamefont
  {Ramos-Prieto}}, \bibinfo {author} {\bibfnamefont {A.}~\bibnamefont
  {Perez-Leija}}, \bibinfo {author} {\bibfnamefont {K.}~\bibnamefont {Busch}},\
  and\ \bibinfo {author} {\bibfnamefont {R.~d.~J.}\ \bibnamefont
  {León-Montiel}},\ }\bibfield  {title} {\bibinfo {title} {Dynamical {Casimir}
  effect in stochastic systems: {Photon} harvesting through noise},\ }\href
  {https://doi.org/10.1103/PhysRevA.96.032501} {\bibfield  {journal} {\bibinfo
  {journal} {Physical Review A}\ }\textbf {\bibinfo {volume} {96}},\ \bibinfo
  {pages} {032501} (\bibinfo {year} {2017})}\BibitemShut {NoStop}%
\bibitem [{\citenamefont {Ma}\ \emph {et~al.}(2019)\citenamefont {Ma},
  \citenamefont {Miao}, \citenamefont {Xiang},\ and\ \citenamefont
  {Zhang}}]{ma_enhanced_2019}%
  \BibitemOpen
  \bibfield  {author} {\bibinfo {author} {\bibfnamefont {S.}~\bibnamefont
  {Ma}}, \bibinfo {author} {\bibfnamefont {H.}~\bibnamefont {Miao}}, \bibinfo
  {author} {\bibfnamefont {Y.}~\bibnamefont {Xiang}},\ and\ \bibinfo {author}
  {\bibfnamefont {S.}~\bibnamefont {Zhang}},\ }\bibfield  {title} {\bibinfo
  {title} {Enhanced {Dynamic} {Casimir} {Effect} in {Temporally} and
  {Spatially} {Modulated} {Josephson} {Transmission} {Line}},\ }\href
  {https://doi.org/10.1002/lpor.201900164} {\bibfield  {journal} {\bibinfo
  {journal} {Laser \& Photonics Reviews}\ }\textbf {\bibinfo {volume} {13}},\
  \bibinfo {pages} {1900164} (\bibinfo {year} {2019})}\BibitemShut {NoStop}%
\bibitem [{\citenamefont {Li}\ \emph {et~al.}(2023)\citenamefont {Li},
  \citenamefont {Lin}, \citenamefont {Wei}, \citenamefont {Liao}, \citenamefont
  {Xu}, \citenamefont {Ke},\ and\ \citenamefont {Yang}}]{li_preparation_2023}%
  \BibitemOpen
  \bibfield  {author} {\bibinfo {author} {\bibfnamefont {N.}~\bibnamefont
  {Li}}, \bibinfo {author} {\bibfnamefont {Z.-J.}\ \bibnamefont {Lin}},
  \bibinfo {author} {\bibfnamefont {M.-S.}\ \bibnamefont {Wei}}, \bibinfo
  {author} {\bibfnamefont {M.-J.}\ \bibnamefont {Liao}}, \bibinfo {author}
  {\bibfnamefont {J.-P.}\ \bibnamefont {Xu}}, \bibinfo {author} {\bibfnamefont
  {S.-H.}\ \bibnamefont {Ke}},\ and\ \bibinfo {author} {\bibfnamefont {Y.-P.}\
  \bibnamefont {Yang}},\ }\bibfield  {title} {\bibinfo {title} {Preparation of
  squeezed light with low average photon number based on dynamic {Casimir}
  effect},\ }\href {https://doi.org/10.1088/1674-1056/acf2ff} {\bibfield
  {journal} {\bibinfo  {journal} {Chinese Physics B}\ }\textbf {\bibinfo
  {volume} {32}},\ \bibinfo {pages} {120301} (\bibinfo {year}
  {2023})}\BibitemShut {NoStop}%
\bibitem [{\citenamefont {Wilson}\ \emph {et~al.}(2011)\citenamefont {Wilson},
  \citenamefont {Johansson}, \citenamefont {Pourkabirian}, \citenamefont
  {Simoen}, \citenamefont {Johansson}, \citenamefont {Duty}, \citenamefont
  {Nori},\ and\ \citenamefont {Delsing}}]{wilson_observation_2011}%
  \BibitemOpen
  \bibfield  {author} {\bibinfo {author} {\bibfnamefont {C.~M.}\ \bibnamefont
  {Wilson}}, \bibinfo {author} {\bibfnamefont {G.}~\bibnamefont {Johansson}},
  \bibinfo {author} {\bibfnamefont {A.}~\bibnamefont {Pourkabirian}}, \bibinfo
  {author} {\bibfnamefont {M.}~\bibnamefont {Simoen}}, \bibinfo {author}
  {\bibfnamefont {J.~R.}\ \bibnamefont {Johansson}}, \bibinfo {author}
  {\bibfnamefont {T.}~\bibnamefont {Duty}}, \bibinfo {author} {\bibfnamefont
  {F.}~\bibnamefont {Nori}},\ and\ \bibinfo {author} {\bibfnamefont
  {P.}~\bibnamefont {Delsing}},\ }\bibfield  {title} {\bibinfo {title}
  {Observation of the dynamical {Casimir} effect in a superconducting
  circuit},\ }\href {https://doi.org/10.1038/nature10561} {\bibfield  {journal}
  {\bibinfo  {journal} {Nature}\ }\textbf {\bibinfo {volume} {479}},\ \bibinfo
  {pages} {376} (\bibinfo {year} {2011})}\BibitemShut {NoStop}%
\bibitem [{\citenamefont {Moussa}\ \emph {et~al.}(2023)\citenamefont {Moussa},
  \citenamefont {Xu}, \citenamefont {Yin}, \citenamefont {Galiffi},
  \citenamefont {Ra’di},\ and\ \citenamefont
  {Alù}}]{moussa_observation_2023}%
  \BibitemOpen
  \bibfield  {author} {\bibinfo {author} {\bibfnamefont {H.}~\bibnamefont
  {Moussa}}, \bibinfo {author} {\bibfnamefont {G.}~\bibnamefont {Xu}}, \bibinfo
  {author} {\bibfnamefont {S.}~\bibnamefont {Yin}}, \bibinfo {author}
  {\bibfnamefont {E.}~\bibnamefont {Galiffi}}, \bibinfo {author} {\bibfnamefont
  {Y.}~\bibnamefont {Ra’di}},\ and\ \bibinfo {author} {\bibfnamefont
  {A.}~\bibnamefont {Alù}},\ }\bibfield  {title} {\bibinfo {title}
  {Observation of temporal reflection and broadband frequency translation at
  photonic time interfaces},\ }\href
  {https://doi.org/10.1038/s41567-023-01975-y} {\bibfield  {journal} {\bibinfo
  {journal} {Nature Physics}\ }\textbf {\bibinfo {volume} {19}},\ \bibinfo
  {pages} {863} (\bibinfo {year} {2023})}\BibitemShut {NoStop}%
\bibitem [{\citenamefont {Jones}\ \emph {et~al.}(2024)\citenamefont {Jones},
  \citenamefont {Kildishev}, \citenamefont {Segev},\ and\ \citenamefont
  {Peroulis}}]{jones_time-reflection_2024}%
  \BibitemOpen
  \bibfield  {author} {\bibinfo {author} {\bibfnamefont {T.~R.}\ \bibnamefont
  {Jones}}, \bibinfo {author} {\bibfnamefont {A.~V.}\ \bibnamefont
  {Kildishev}}, \bibinfo {author} {\bibfnamefont {M.}~\bibnamefont {Segev}},\
  and\ \bibinfo {author} {\bibfnamefont {D.}~\bibnamefont {Peroulis}},\
  }\bibfield  {title} {\bibinfo {title} {Time-reflection of microwaves by a
  fast optically-controlled time-boundary},\ }\href
  {https://doi.org/10.1038/s41467-024-51171-6} {\bibfield  {journal} {\bibinfo
  {journal} {Nature Communications}\ }\textbf {\bibinfo {volume} {15}},\
  \bibinfo {pages} {6786} (\bibinfo {year} {2024})}\BibitemShut {NoStop}%
\bibitem [{\citenamefont {Caspani}\ \emph {et~al.}(2016)\citenamefont
  {Caspani}, \citenamefont {Kaipurath}, \citenamefont {Clerici}, \citenamefont
  {Ferrera}, \citenamefont {Roger}, \citenamefont {Kim}, \citenamefont
  {Kinsey}, \citenamefont {Pietrzyk}, \citenamefont {Di~Falco}, \citenamefont
  {Shalaev}, \citenamefont {Boltasseva},\ and\ \citenamefont
  {Faccio}}]{caspani_enhanced_2016}%
  \BibitemOpen
  \bibfield  {author} {\bibinfo {author} {\bibfnamefont {L.}~\bibnamefont
  {Caspani}}, \bibinfo {author} {\bibfnamefont {R.}~\bibnamefont {Kaipurath}},
  \bibinfo {author} {\bibfnamefont {M.}~\bibnamefont {Clerici}}, \bibinfo
  {author} {\bibfnamefont {M.}~\bibnamefont {Ferrera}}, \bibinfo {author}
  {\bibfnamefont {T.}~\bibnamefont {Roger}}, \bibinfo {author} {\bibfnamefont
  {J.}~\bibnamefont {Kim}}, \bibinfo {author} {\bibfnamefont {N.}~\bibnamefont
  {Kinsey}}, \bibinfo {author} {\bibfnamefont {M.}~\bibnamefont {Pietrzyk}},
  \bibinfo {author} {\bibfnamefont {A.}~\bibnamefont {Di~Falco}}, \bibinfo
  {author} {\bibfnamefont {V.}~\bibnamefont {Shalaev}}, \bibinfo {author}
  {\bibfnamefont {A.}~\bibnamefont {Boltasseva}},\ and\ \bibinfo {author}
  {\bibfnamefont {D.}~\bibnamefont {Faccio}},\ }\bibfield  {title} {\bibinfo
  {title} {Enhanced {Nonlinear} {Refractive} {Index} in
  ${\varepsilon}$-{Near}-{Zero} {Materials}},\ }\href
  {https://doi.org/10.1103/PhysRevLett.116.233901} {\bibfield  {journal}
  {\bibinfo  {journal} {Physical Review Letters}\ }\textbf {\bibinfo {volume}
  {116}},\ \bibinfo {pages} {233901} (\bibinfo {year} {2016})}\BibitemShut
  {NoStop}%
\bibitem [{\citenamefont {Zhou}\ \emph {et~al.}(2020)\citenamefont {Zhou},
  \citenamefont {Alam}, \citenamefont {Karimi}, \citenamefont {Upham},
  \citenamefont {Reshef}, \citenamefont {Liu}, \citenamefont {Willner},\ and\
  \citenamefont {Boyd}}]{zhou_broadband_2020}%
  \BibitemOpen
  \bibfield  {author} {\bibinfo {author} {\bibfnamefont {Y.}~\bibnamefont
  {Zhou}}, \bibinfo {author} {\bibfnamefont {M.~Z.}\ \bibnamefont {Alam}},
  \bibinfo {author} {\bibfnamefont {M.}~\bibnamefont {Karimi}}, \bibinfo
  {author} {\bibfnamefont {J.}~\bibnamefont {Upham}}, \bibinfo {author}
  {\bibfnamefont {O.}~\bibnamefont {Reshef}}, \bibinfo {author} {\bibfnamefont
  {C.}~\bibnamefont {Liu}}, \bibinfo {author} {\bibfnamefont {A.~E.}\
  \bibnamefont {Willner}},\ and\ \bibinfo {author} {\bibfnamefont {R.~W.}\
  \bibnamefont {Boyd}},\ }\bibfield  {title} {\bibinfo {title} {Broadband
  frequency translation through time refraction in an epsilon-near-zero
  material},\ }\href {https://doi.org/10.1038/s41467-020-15682-2} {\bibfield
  {journal} {\bibinfo  {journal} {Nature Communications}\ }\textbf {\bibinfo
  {volume} {11}},\ \bibinfo {pages} {2180} (\bibinfo {year}
  {2020})}\BibitemShut {NoStop}%
\bibitem [{\citenamefont {Bohn}\ \emph {et~al.}(2021)\citenamefont {Bohn},
  \citenamefont {Luk}, \citenamefont {Horsley},\ and\ \citenamefont
  {Hendry}}]{bohn_spatiotemporal_2021}%
  \BibitemOpen
  \bibfield  {author} {\bibinfo {author} {\bibfnamefont {J.}~\bibnamefont
  {Bohn}}, \bibinfo {author} {\bibfnamefont {T.~S.}\ \bibnamefont {Luk}},
  \bibinfo {author} {\bibfnamefont {S.}~\bibnamefont {Horsley}},\ and\ \bibinfo
  {author} {\bibfnamefont {E.}~\bibnamefont {Hendry}},\ }\bibfield  {title}
  {\bibinfo {title} {Spatiotemporal refraction of light in an epsilon-near-zero
  indium tin oxide layer: frequency shifting effects arising from interfaces},\
  }\href {https://doi.org/10.1364/OPTICA.436324} {\bibfield  {journal}
  {\bibinfo  {journal} {Optica}\ }\textbf {\bibinfo {volume} {8}},\ \bibinfo
  {pages} {1532} (\bibinfo {year} {2021})}\BibitemShut {NoStop}%
\bibitem [{\citenamefont {Tirole}\ \emph {et~al.}(2023)\citenamefont {Tirole},
  \citenamefont {Vezzoli}, \citenamefont {Galiffi}, \citenamefont {Robertson},
  \citenamefont {Maurice}, \citenamefont {Tilmann}, \citenamefont {Maier},
  \citenamefont {Pendry},\ and\ \citenamefont
  {Sapienza}}]{tirole_double-slit_2023}%
  \BibitemOpen
  \bibfield  {author} {\bibinfo {author} {\bibfnamefont {R.}~\bibnamefont
  {Tirole}}, \bibinfo {author} {\bibfnamefont {S.}~\bibnamefont {Vezzoli}},
  \bibinfo {author} {\bibfnamefont {E.}~\bibnamefont {Galiffi}}, \bibinfo
  {author} {\bibfnamefont {I.}~\bibnamefont {Robertson}}, \bibinfo {author}
  {\bibfnamefont {D.}~\bibnamefont {Maurice}}, \bibinfo {author} {\bibfnamefont
  {B.}~\bibnamefont {Tilmann}}, \bibinfo {author} {\bibfnamefont {S.~A.}\
  \bibnamefont {Maier}}, \bibinfo {author} {\bibfnamefont {J.~B.}\ \bibnamefont
  {Pendry}},\ and\ \bibinfo {author} {\bibfnamefont {R.}~\bibnamefont
  {Sapienza}},\ }\bibfield  {title} {\bibinfo {title} {Double-slit time
  diffraction at optical frequencies},\ }\href
  {https://doi.org/10.1038/s41567-023-01993-w} {\bibfield  {journal} {\bibinfo
  {journal} {Nature Physics}\ }\textbf {\bibinfo {volume} {19}},\ \bibinfo
  {pages} {999} (\bibinfo {year} {2023})}\BibitemShut {NoStop}%
\bibitem [{\citenamefont {Galiffi}\ \emph {et~al.}(2024)\citenamefont
  {Galiffi}, \citenamefont {Harwood}, \citenamefont {Vezzoli}, \citenamefont
  {Tirole}, \citenamefont {Alù},\ and\ \citenamefont
  {Sapienza}}]{galiffi_optical_2024}%
  \BibitemOpen
  \bibfield  {author} {\bibinfo {author} {\bibfnamefont {E.}~\bibnamefont
  {Galiffi}}, \bibinfo {author} {\bibfnamefont {A.~C.}\ \bibnamefont
  {Harwood}}, \bibinfo {author} {\bibfnamefont {S.}~\bibnamefont {Vezzoli}},
  \bibinfo {author} {\bibfnamefont {R.}~\bibnamefont {Tirole}}, \bibinfo
  {author} {\bibfnamefont {A.}~\bibnamefont {Alù}},\ and\ \bibinfo {author}
  {\bibfnamefont {R.}~\bibnamefont {Sapienza}},\ }\href
  {https://doi.org/10.48550/arXiv.2410.16426} {\bibinfo {title} {Optical
  coherent perfect absorption and amplification in a time-varying medium}}
  (\bibinfo {year} {2024}),\ \bibinfo {note} {arXiv:2410.16426}\BibitemShut
  {NoStop}%
\bibitem [{\citenamefont {Pacheco-Peña}\ and\ \citenamefont
  {Engheta}(2020)}]{pacheco-pena_antireflection_2020}%
  \BibitemOpen
  \bibfield  {author} {\bibinfo {author} {\bibfnamefont {V.}~\bibnamefont
  {Pacheco-Peña}}\ and\ \bibinfo {author} {\bibfnamefont {N.}~\bibnamefont
  {Engheta}},\ }\bibfield  {title} {\bibinfo {title} {Antireflection temporal
  coatings},\ }\href@noop {} {\bibfield  {journal} {\bibinfo  {journal}
  {Optica}\ }\textbf {\bibinfo {volume} {7}},\ \bibinfo {pages} {323} (\bibinfo
  {year} {2020})}\BibitemShut {NoStop}%
\bibitem [{\citenamefont {Liberal}\ \emph {et~al.}(2023)\citenamefont
  {Liberal}, \citenamefont {Vázquez‐Lozano},\ and\ \citenamefont
  {Pacheco‐Peña}}]{liberal_quantum_2023}%
  \BibitemOpen
  \bibfield  {author} {\bibinfo {author} {\bibfnamefont {I.}~\bibnamefont
  {Liberal}}, \bibinfo {author} {\bibfnamefont {J.~E.}\ \bibnamefont
  {Vázquez‐Lozano}},\ and\ \bibinfo {author} {\bibfnamefont
  {V.}~\bibnamefont {Pacheco‐Peña}},\ }\bibfield  {title} {\bibinfo {title}
  {Quantum {Antireflection} {Temporal} {Coatings}: {Quantum} {State}
  {Frequency} {Shifting} and {Inhibited} {Thermal} {Noise} {Amplification}},\
  }\href {https://doi.org/10.1002/lpor.202200720} {\bibfield  {journal}
  {\bibinfo  {journal} {Laser \& Photonics Reviews}\ }\textbf {\bibinfo
  {volume} {17}},\ \bibinfo {pages} {2200720} (\bibinfo {year}
  {2023})}\BibitemShut {NoStop}%
\bibitem [{\citenamefont {Dikopoltsev}\ \emph {et~al.}(2022)\citenamefont
  {Dikopoltsev}, \citenamefont {Sharabi}, \citenamefont {Lyubarov},
  \citenamefont {Lumer}, \citenamefont {Tsesses}, \citenamefont {Lustig},
  \citenamefont {Kaminer},\ and\ \citenamefont
  {Segev}}]{dikopoltsev_light_2022}%
  \BibitemOpen
  \bibfield  {author} {\bibinfo {author} {\bibfnamefont {A.}~\bibnamefont
  {Dikopoltsev}}, \bibinfo {author} {\bibfnamefont {Y.}~\bibnamefont
  {Sharabi}}, \bibinfo {author} {\bibfnamefont {M.}~\bibnamefont {Lyubarov}},
  \bibinfo {author} {\bibfnamefont {Y.}~\bibnamefont {Lumer}}, \bibinfo
  {author} {\bibfnamefont {S.}~\bibnamefont {Tsesses}}, \bibinfo {author}
  {\bibfnamefont {E.}~\bibnamefont {Lustig}}, \bibinfo {author} {\bibfnamefont
  {I.}~\bibnamefont {Kaminer}},\ and\ \bibinfo {author} {\bibfnamefont
  {M.}~\bibnamefont {Segev}},\ }\bibfield  {title} {\bibinfo {title} {Light
  emission by free electrons in photonic time-crystals},\ }\href
  {https://doi.org/10.1073/pnas.2119705119} {\bibfield  {journal} {\bibinfo
  {journal} {Proceedings of the National Academy of Sciences}\ }\textbf
  {\bibinfo {volume} {119}},\ \bibinfo {pages} {e2119705119} (\bibinfo {year}
  {2022})}\BibitemShut {NoStop}%
\bibitem [{\citenamefont {Lyubarov}\ \emph {et~al.}(2022)\citenamefont
  {Lyubarov}, \citenamefont {Lumer}, \citenamefont {Dikopoltsev}, \citenamefont
  {Lustig}, \citenamefont {Sharabi},\ and\ \citenamefont
  {Segev}}]{lyubarov_amplified_2022}%
  \BibitemOpen
  \bibfield  {author} {\bibinfo {author} {\bibfnamefont {M.}~\bibnamefont
  {Lyubarov}}, \bibinfo {author} {\bibfnamefont {Y.}~\bibnamefont {Lumer}},
  \bibinfo {author} {\bibfnamefont {A.}~\bibnamefont {Dikopoltsev}}, \bibinfo
  {author} {\bibfnamefont {E.}~\bibnamefont {Lustig}}, \bibinfo {author}
  {\bibfnamefont {Y.}~\bibnamefont {Sharabi}},\ and\ \bibinfo {author}
  {\bibfnamefont {M.}~\bibnamefont {Segev}},\ }\bibfield  {title} {\bibinfo
  {title} {Amplified emission and lasing in photonic time crystals},\ }\href
  {https://doi.org/10.1126/science.abo3324} {\bibfield  {journal} {\bibinfo
  {journal} {Science}\ }\textbf {\bibinfo {volume} {377}},\ \bibinfo {pages}
  {425} (\bibinfo {year} {2022})}\BibitemShut {NoStop}%
\bibitem [{\citenamefont {Lyubarov}\ \emph {et~al.}(2024)\citenamefont
  {Lyubarov}, \citenamefont {Dikopoltsev}, \citenamefont {Segal}, \citenamefont
  {Plotnik},\ and\ \citenamefont {Segev}}]{lyubarov_controlling_2024}%
  \BibitemOpen
  \bibfield  {author} {\bibinfo {author} {\bibfnamefont {M.}~\bibnamefont
  {Lyubarov}}, \bibinfo {author} {\bibfnamefont {A.}~\bibnamefont
  {Dikopoltsev}}, \bibinfo {author} {\bibfnamefont {O.}~\bibnamefont {Segal}},
  \bibinfo {author} {\bibfnamefont {Y.}~\bibnamefont {Plotnik}},\ and\ \bibinfo
  {author} {\bibfnamefont {M.}~\bibnamefont {Segev}},\ }\bibfield  {title}
  {\bibinfo {title} {Controlling spontaneous emission through the preparation
  of a photonic time-crystal},\ }\href {https://doi.org/10.1364/OE.539636}
  {\bibfield  {journal} {\bibinfo  {journal} {Optics Express}\ }\textbf
  {\bibinfo {volume} {32}},\ \bibinfo {pages} {39734} (\bibinfo {year}
  {2024})}\BibitemShut {NoStop}%
\bibitem [{\citenamefont {Vázquez-Lozano}\ and\ \citenamefont
  {Liberal}(2023)}]{vazquez-lozano_shaping_2023}%
  \BibitemOpen
  \bibfield  {author} {\bibinfo {author} {\bibfnamefont {J.~E.}\ \bibnamefont
  {Vázquez-Lozano}}\ and\ \bibinfo {author} {\bibfnamefont {I.}~\bibnamefont
  {Liberal}},\ }\bibfield  {title} {\bibinfo {title} {Shaping the quantum
  vacuum with anisotropic temporal boundaries},\ }\href
  {https://doi.org/10.1515/nanoph-2022-0491} {\bibfield  {journal} {\bibinfo
  {journal} {Nanophotonics}\ }\textbf {\bibinfo {volume} {12}},\ \bibinfo
  {pages} {539} (\bibinfo {year} {2023})}\BibitemShut {NoStop}%
\bibitem [{\citenamefont {Ganfornina-Andrades}\ \emph
  {et~al.}(2024)\citenamefont {Ganfornina-Andrades}, \citenamefont
  {Vázquez-Lozano},\ and\ \citenamefont
  {Liberal}}]{ganfornina-andrades_quantum_2023}%
  \BibitemOpen
  \bibfield  {author} {\bibinfo {author} {\bibfnamefont {A.}~\bibnamefont
  {Ganfornina-Andrades}}, \bibinfo {author} {\bibfnamefont {J.~E.}\
  \bibnamefont {Vázquez-Lozano}},\ and\ \bibinfo {author} {\bibfnamefont
  {I.}~\bibnamefont {Liberal}},\ }\bibfield  {title} {\bibinfo {title} {Quantum
  vacuum amplification in time-varying media with arbitrary temporal
  profiles},\ }\href@noop {} {\bibfield  {journal} {\bibinfo  {journal}
  {Physical Review Research}\ }\textbf {\bibinfo {volume} {6}},\ \bibinfo
  {pages} {043320} (\bibinfo {year} {2024})}\BibitemShut {NoStop}%
\bibitem [{\citenamefont {Mendonça}\ \emph {et~al.}(2003)\citenamefont
  {Mendonça}, \citenamefont {Martins},\ and\ \citenamefont
  {Guerreiro}}]{mendonca_temporal_2003}%
  \BibitemOpen
  \bibfield  {author} {\bibinfo {author} {\bibfnamefont {J.~T.}\ \bibnamefont
  {Mendonça}}, \bibinfo {author} {\bibfnamefont {A.~M.}\ \bibnamefont
  {Martins}},\ and\ \bibinfo {author} {\bibfnamefont {A.}~\bibnamefont
  {Guerreiro}},\ }\bibfield  {title} {\bibinfo {title} {Temporal beam splitter
  and temporal interference},\ }\href
  {https://doi.org/10.1103/PhysRevA.68.043801} {\bibfield  {journal} {\bibinfo
  {journal} {Physical Review A}\ }\textbf {\bibinfo {volume} {68}},\ \bibinfo
  {pages} {043801} (\bibinfo {year} {2003})}\BibitemShut {NoStop}%
\bibitem [{\citenamefont {Mendonça}\ and\ \citenamefont
  {Guerreiro}(2005)}]{mendonca_time_2005}%
  \BibitemOpen
  \bibfield  {author} {\bibinfo {author} {\bibfnamefont {J.~T.}\ \bibnamefont
  {Mendonça}}\ and\ \bibinfo {author} {\bibfnamefont {A.}~\bibnamefont
  {Guerreiro}},\ }\bibfield  {title} {\bibinfo {title} {Time refraction and the
  quantum properties of vacuum},\ }\href
  {https://doi.org/10.1103/PhysRevA.72.063805} {\bibfield  {journal} {\bibinfo
  {journal} {Physical Review A}\ }\textbf {\bibinfo {volume} {72}},\ \bibinfo
  {pages} {063805} (\bibinfo {year} {2005})}\BibitemShut {NoStop}%
\bibitem [{\citenamefont {Mirmoosa}\ \emph {et~al.}(2025)\citenamefont
  {Mirmoosa}, \citenamefont {Setälä},\ and\ \citenamefont
  {Norrman}}]{mirmoosa_quantum_2025}%
  \BibitemOpen
  \bibfield  {author} {\bibinfo {author} {\bibfnamefont {M.~S.}\ \bibnamefont
  {Mirmoosa}}, \bibinfo {author} {\bibfnamefont {T.}~\bibnamefont {Setälä}},\
  and\ \bibinfo {author} {\bibfnamefont {A.}~\bibnamefont {Norrman}},\
  }\bibfield  {title} {\bibinfo {title} {Quantum state engineering and photon
  statistics at electromagnetic time interfaces},\ }\href
  {https://doi.org/10.1103/PhysRevResearch.7.013120} {\bibfield  {journal}
  {\bibinfo  {journal} {Physical Review Research}\ }\textbf {\bibinfo {volume}
  {7}},\ \bibinfo {pages} {013120} (\bibinfo {year} {2025})}\BibitemShut
  {NoStop}%
\bibitem [{\citenamefont {Hillery}(2009)}]{hillery_introduction_2009}%
  \BibitemOpen
  \bibfield  {author} {\bibinfo {author} {\bibfnamefont {M.}~\bibnamefont
  {Hillery}},\ }\bibfield  {title} {\bibinfo {title} {An introduction to the
  quantum theory of nonlinear optics},\ }\href@noop {} {\bibfield  {journal}
  {\bibinfo  {journal} {Acta Physica Slovaca}\ }\textbf {\bibinfo {volume}
  {59}},\ \bibinfo {pages} {1} (\bibinfo {year} {2009})}\BibitemShut {NoStop}%
\bibitem [{\citenamefont {Cohen-Tannoudji}\ \emph {et~al.}(2024)\citenamefont
  {Cohen-Tannoudji}, \citenamefont {Dupont-Roc},\ and\ \citenamefont
  {Grynberg}}]{cohen-tannoudji_photons_2024}%
  \BibitemOpen
  \bibfield  {author} {\bibinfo {author} {\bibfnamefont {C.}~\bibnamefont
  {Cohen-Tannoudji}}, \bibinfo {author} {\bibfnamefont {J.}~\bibnamefont
  {Dupont-Roc}},\ and\ \bibinfo {author} {\bibfnamefont {G.}~\bibnamefont
  {Grynberg}},\ }\href@noop {} {\emph {\bibinfo {title} {Photons and {Atoms}:
  {Introduction} to {Quantum} {Electrodynamics}}}}\ (\bibinfo  {publisher}
  {John Wiley \& Sons},\ \bibinfo {year} {2024})\BibitemShut {NoStop}%
\bibitem [{\citenamefont {Loudon}(2000)}]{loudon_quantum_2000}%
  \BibitemOpen
  \bibfield  {author} {\bibinfo {author} {\bibfnamefont {R.}~\bibnamefont
  {Loudon}},\ }\href@noop {} {\emph {\bibinfo {title} {The {Quantum} {Theory}
  of {Light}}}}\ (\bibinfo  {publisher} {OUP Oxford},\ \bibinfo {year}
  {2000})\BibitemShut {NoStop}%
\bibitem [{Sup()}]{Supplementary_material}%
  \BibitemOpen
  \href@noop {} {\bibinfo {title} {See supplemental material at [url will be
  inserted by publisher] for detailed derivations.}}\BibitemShut {Stop}%
\bibitem [{\citenamefont {Stevens}\ and\ \citenamefont
  {Caloz}(2024)}]{stevens_photon_2024}%
  \BibitemOpen
  \bibfield  {author} {\bibinfo {author} {\bibfnamefont {A.}~\bibnamefont
  {Stevens}}\ and\ \bibinfo {author} {\bibfnamefont {C.}~\bibnamefont
  {Caloz}},\ }\bibfield  {title} {\bibinfo {title} {Photon transitions in
  arbitrary time-varying metamaterials},\ }in\ \href
  {https://doi.org/10.1109/Metamaterials62190.2024.10703217} {\emph {\bibinfo
  {booktitle} {2024 Eighteenth International Congress on Artificial Materials
  for Novel Wave Phenomena (Metamaterials)}}}\ (\bibinfo {year} {2024})\ pp.\
  \bibinfo {pages} {1--3}\BibitemShut {NoStop}%
\bibitem [{\citenamefont {Morgenthaler}(1958)}]{morgenthaler_velocity_1958}%
  \BibitemOpen
  \bibfield  {author} {\bibinfo {author} {\bibfnamefont {F.}~\bibnamefont
  {Morgenthaler}},\ }\bibfield  {title} {\bibinfo {title} {Velocity
  {Modulation} of {Electromagnetic} {Waves}},\ }\href
  {https://doi.org/10.1109/TMTT.1958.1124533} {\bibfield  {journal} {\bibinfo
  {journal} {IRE Transactions on Microwave Theory and Techniques}\ }\textbf
  {\bibinfo {volume} {6}},\ \bibinfo {pages} {167} (\bibinfo {year}
  {1958})}\BibitemShut {NoStop}%
\bibitem [{\citenamefont {Caloz}\ and\ \citenamefont
  {Deck-Leger}(2019)}]{caloz_spacetime_2019}%
  \BibitemOpen
  \bibfield  {author} {\bibinfo {author} {\bibfnamefont {C.}~\bibnamefont
  {Caloz}}\ and\ \bibinfo {author} {\bibfnamefont {Z.-L.}\ \bibnamefont
  {Deck-Leger}},\ }\bibfield  {title} {\bibinfo {title} {Spacetime
  metamaterials—{Part} {II}: {Theory} and applications},\ }\href
  {https://ieeexplore.ieee.org/abstract/document/8858032/} {\bibfield
  {journal} {\bibinfo  {journal} {IEEE Transactions on Antennas and
  Propagation}\ }\textbf {\bibinfo {volume} {68}},\ \bibinfo {pages} {1583}
  (\bibinfo {year} {2019})}\BibitemShut {NoStop}%
\bibitem [{\citenamefont {Koutserimpas}\ and\ \citenamefont
  {Fleury}(2020)}]{koutserimpas_electromagnetic_2020}%
  \BibitemOpen
  \bibfield  {author} {\bibinfo {author} {\bibfnamefont {T.~T.}\ \bibnamefont
  {Koutserimpas}}\ and\ \bibinfo {author} {\bibfnamefont {R.}~\bibnamefont
  {Fleury}},\ }\bibfield  {title} {\bibinfo {title} {Electromagnetic {Fields}
  in a {Time}-{Varying} {Medium}: {Exceptional} {Points} and {Operator}
  {Symmetries}},\ }\href {https://doi.org/10.1109/TAP.2020.2996822} {\bibfield
  {journal} {\bibinfo  {journal} {IEEE Transactions on Antennas and
  Propagation}\ }\textbf {\bibinfo {volume} {68}},\ \bibinfo {pages} {6717}
  (\bibinfo {year} {2020})}\BibitemShut {NoStop}%
\bibitem [{\citenamefont {Ortega-Gomez}\ \emph {et~al.}(2023)\citenamefont
  {Ortega-Gomez}, \citenamefont {Lobet}, \citenamefont {Vázquez-Lozano},\ and\
  \citenamefont {Liberal}}]{ortega-gomez_tutorial_2023}%
  \BibitemOpen
  \bibfield  {author} {\bibinfo {author} {\bibfnamefont {A.}~\bibnamefont
  {Ortega-Gomez}}, \bibinfo {author} {\bibfnamefont {M.}~\bibnamefont {Lobet}},
  \bibinfo {author} {\bibfnamefont {J.~E.}\ \bibnamefont {Vázquez-Lozano}},\
  and\ \bibinfo {author} {\bibfnamefont {I.}~\bibnamefont {Liberal}},\
  }\bibfield  {title} {\bibinfo {title} {Tutorial on the conservation of
  momentum in photonic time-varying media [{Invited}]},\ }\href
  {https://doi.org/10.1364/OME.485540} {\bibfield  {journal} {\bibinfo
  {journal} {Optical Materials Express}\ }\textbf {\bibinfo {volume} {13}},\
  \bibinfo {pages} {1598} (\bibinfo {year} {2023})}\BibitemShut {NoStop}%
\bibitem [{\citenamefont {Galiffi}\ \emph {et~al.}(2022)\citenamefont
  {Galiffi}, \citenamefont {Tirole}, \citenamefont {Yin}, \citenamefont {Li},
  \citenamefont {Vezzoli}, \citenamefont {Huidobro}, \citenamefont
  {Silveirinha}, \citenamefont {Sapienza}, \citenamefont {Alù},\ and\
  \citenamefont {Pendry}}]{galiffi_photonics_2022}%
  \BibitemOpen
  \bibfield  {author} {\bibinfo {author} {\bibfnamefont {E.}~\bibnamefont
  {Galiffi}}, \bibinfo {author} {\bibfnamefont {R.}~\bibnamefont {Tirole}},
  \bibinfo {author} {\bibfnamefont {S.}~\bibnamefont {Yin}}, \bibinfo {author}
  {\bibfnamefont {H.}~\bibnamefont {Li}}, \bibinfo {author} {\bibfnamefont
  {S.}~\bibnamefont {Vezzoli}}, \bibinfo {author} {\bibfnamefont {P.~A.}\
  \bibnamefont {Huidobro}}, \bibinfo {author} {\bibfnamefont {M.~G.}\
  \bibnamefont {Silveirinha}}, \bibinfo {author} {\bibfnamefont
  {R.}~\bibnamefont {Sapienza}}, \bibinfo {author} {\bibfnamefont
  {A.}~\bibnamefont {Alù}},\ and\ \bibinfo {author} {\bibfnamefont {J.~B.}\
  \bibnamefont {Pendry}},\ }\bibfield  {title} {\bibinfo {title} {Photonics of
  time-varying media},\ }\href {https://doi.org/10.1117/1.AP.4.1.014002}
  {\bibfield  {journal} {\bibinfo  {journal} {Advanced Photonics}\ }\textbf
  {\bibinfo {volume} {4}},\ \bibinfo {pages} {014002} (\bibinfo {year}
  {2022})}\BibitemShut {NoStop}%
\bibitem [{\citenamefont {Boyd}(2008)}]{BoydRobertW.2008NO(E}%
  \BibitemOpen
  \bibfield  {author} {\bibinfo {author} {\bibfnamefont {R.~W.}\ \bibnamefont
  {Boyd}},\ }\href@noop {} {\emph {\bibinfo {title} {Nonlinear Optics (3rd
  Edition)}}},\ \bibinfo {edition} {2nd}\ ed.\ (\bibinfo  {publisher}
  {Elsevier},\ \bibinfo {address} {Chantilly},\ \bibinfo {year}
  {2008})\BibitemShut {NoStop}%
\bibitem [{\citenamefont {Scully}\ and\ \citenamefont
  {Zubairy}(1997)}]{scully_quantum_1997}%
  \BibitemOpen
  \bibfield  {author} {\bibinfo {author} {\bibfnamefont {M.~O.}\ \bibnamefont
  {Scully}}\ and\ \bibinfo {author} {\bibfnamefont {M.~S.}\ \bibnamefont
  {Zubairy}},\ }\href@noop {} {\emph {\bibinfo {title} {Quantum {Optics}}}}\
  (\bibinfo  {publisher} {Cambridge University Press},\ \bibinfo {year}
  {1997})\BibitemShut {NoStop}%
\bibitem [{\citenamefont {Dyson}(1949)}]{dyson_s_1949}%
  \BibitemOpen
  \bibfield  {author} {\bibinfo {author} {\bibfnamefont {F.~J.}\ \bibnamefont
  {Dyson}},\ }\bibfield  {title} {\bibinfo {title} {The {$S$} {Matrix} in
  {Quantum} {Electrodynamics}},\ }\href
  {https://doi.org/10.1103/PhysRev.75.1736} {\bibfield  {journal} {\bibinfo
  {journal} {Physical Review}\ }\textbf {\bibinfo {volume} {75}},\ \bibinfo
  {pages} {1736} (\bibinfo {year} {1949})}\BibitemShut {NoStop}%
\bibitem [{\citenamefont {Sakurai}\ and\ \citenamefont
  {Napolitano}(2020)}]{sakurai_modern_2020}%
  \BibitemOpen
  \bibfield  {author} {\bibinfo {author} {\bibfnamefont {J.~J.}\ \bibnamefont
  {Sakurai}}\ and\ \bibinfo {author} {\bibfnamefont {J.}~\bibnamefont
  {Napolitano}},\ }\href@noop {} {\emph {\bibinfo {title} {Modern {Quantum}
  {Mechanics}}}}\ (\bibinfo  {publisher} {Cambridge University Press},\
  \bibinfo {year} {2020})\BibitemShut {NoStop}%
\bibitem [{\citenamefont {Bennemann}\ and\ \citenamefont
  {Ketterson}(2008)}]{bennemann_superconductivity_2008}%
  \BibitemOpen
  \bibfield  {author} {\bibinfo {author} {\bibfnamefont {K.-H.}\ \bibnamefont
  {Bennemann}}\ and\ \bibinfo {author} {\bibfnamefont {J.~B.}\ \bibnamefont
  {Ketterson}},\ }\href@noop {} {\emph {\bibinfo {title} {Superconductivity:
  {Volume} 1: {Conventional} and {Unconventional} {Superconductors} {Volume} 2:
  {Novel} {Superconductors}}}}\ (\bibinfo  {publisher} {Springer Science \&
  Business Media},\ \bibinfo {year} {2008})\BibitemShut {NoStop}%
\bibitem [{\citenamefont {Bogoljubov}\ \emph {et~al.}(1958)\citenamefont
  {Bogoljubov}, \citenamefont {Tolmachov},\ and\ \citenamefont
  {{\v{S}}irkov}}]{bogoljubov1958new}%
  \BibitemOpen
  \bibfield  {author} {\bibinfo {author} {\bibfnamefont {N.~N.}\ \bibnamefont
  {Bogoljubov}}, \bibinfo {author} {\bibfnamefont {V.~V.}\ \bibnamefont
  {Tolmachov}},\ and\ \bibinfo {author} {\bibfnamefont {D.}~\bibnamefont
  {{\v{S}}irkov}},\ }\bibfield  {title} {\bibinfo {title} {A new method in the
  theory of superconductivity},\ }\href@noop {} {\bibfield  {journal} {\bibinfo
   {journal} {Fortschritte der physik}\ }\textbf {\bibinfo {volume} {6}},\
  \bibinfo {pages} {605} (\bibinfo {year} {1958})}\BibitemShut {NoStop}%
\bibitem [{\citenamefont {Castaldi}\ \emph {et~al.}(2021)\citenamefont
  {Castaldi}, \citenamefont {Pacheco-Pe{\~n}a}, \citenamefont {Moccia},
  \citenamefont {Engheta},\ and\ \citenamefont
  {Galdi}}]{castaldi2021exploiting}%
  \BibitemOpen
  \bibfield  {author} {\bibinfo {author} {\bibfnamefont {G.}~\bibnamefont
  {Castaldi}}, \bibinfo {author} {\bibfnamefont {V.}~\bibnamefont
  {Pacheco-Pe{\~n}a}}, \bibinfo {author} {\bibfnamefont {M.}~\bibnamefont
  {Moccia}}, \bibinfo {author} {\bibfnamefont {N.}~\bibnamefont {Engheta}},\
  and\ \bibinfo {author} {\bibfnamefont {V.}~\bibnamefont {Galdi}},\ }\bibfield
   {title} {\bibinfo {title} {Exploiting space-time duality in the synthesis of
  impedance transformers via temporal metamaterials},\ }\href@noop {}
  {\bibfield  {journal} {\bibinfo  {journal} {Nanophotonics}\ }\textbf
  {\bibinfo {volume} {10}},\ \bibinfo {pages} {3687} (\bibinfo {year}
  {2021})}\BibitemShut {NoStop}%
\bibitem [{\citenamefont {Sloan}\ \emph {et~al.}(2021)\citenamefont {Sloan},
  \citenamefont {Rivera}, \citenamefont {Joannopoulos},\ and\ \citenamefont
  {Soljačić}}]{sloan_casimir_2021}%
  \BibitemOpen
  \bibfield  {author} {\bibinfo {author} {\bibfnamefont {J.}~\bibnamefont
  {Sloan}}, \bibinfo {author} {\bibfnamefont {N.}~\bibnamefont {Rivera}},
  \bibinfo {author} {\bibfnamefont {J.~D.}\ \bibnamefont {Joannopoulos}},\ and\
  \bibinfo {author} {\bibfnamefont {M.}~\bibnamefont {Soljačić}},\ }\bibfield
   {title} {\bibinfo {title} {Casimir {Light} in {Dispersive}
  {Nanophotonics}},\ }\href@noop {} {\bibfield  {journal} {\bibinfo  {journal}
  {Physical Review Letters}\ }\textbf {\bibinfo {volume} {127}},\ \bibinfo
  {pages} {053603} (\bibinfo {year} {2021})}\BibitemShut {NoStop}%
\bibitem [{\citenamefont {Sloan}\ \emph {et~al.}(2024)\citenamefont {Sloan},
  \citenamefont {Rivera}, \citenamefont {Joannopoulos},\ and\ \citenamefont
  {Soljacic}}]{sloan_optical_2024}%
  \BibitemOpen
  \bibfield  {author} {\bibinfo {author} {\bibfnamefont {J.}~\bibnamefont
  {Sloan}}, \bibinfo {author} {\bibfnamefont {N.}~\bibnamefont {Rivera}},
  \bibinfo {author} {\bibfnamefont {J.~D.}\ \bibnamefont {Joannopoulos}},\ and\
  \bibinfo {author} {\bibfnamefont {M.}~\bibnamefont {Soljacic}},\ }\bibfield
  {title} {\bibinfo {title} {Optical {Properties} of {Dispersive}
  {Time}-{Dependent} {Materials}},\ }\href@noop {} {\bibfield  {journal}
  {\bibinfo  {journal} {ACS Photonics}\ }\textbf {\bibinfo {volume} {11}},\
  \bibinfo {pages} {950} (\bibinfo {year} {2024})}\BibitemShut {NoStop}%
\bibitem [{\citenamefont {Horsley}\ and\ \citenamefont
  {Baker}(2025)}]{horsley_macroscopic_2024}%
  \BibitemOpen
  \bibfield  {author} {\bibinfo {author} {\bibfnamefont {S.~A.~R.}\
  \bibnamefont {Horsley}}\ and\ \bibinfo {author} {\bibfnamefont {R.~K.}\
  \bibnamefont {Baker}},\ }\bibfield  {title} {\bibinfo {title} {Macroscopic
  {QED} and noise currents in time-varying media},\ }\href
  {https://doi.org/10.1103/PhysRevA.111.053511} {\bibfield  {journal} {\bibinfo
   {journal} {Physical Review A}\ }\textbf {\bibinfo {volume} {111}},\ \bibinfo
  {pages} {053511} (\bibinfo {year} {2025})}\BibitemShut {NoStop}%
\bibitem [{\citenamefont {Koutserimpas}\ and\ \citenamefont
  {Monticone}(2024)}]{koutserimpas_time-varying_2024}%
  \BibitemOpen
  \bibfield  {author} {\bibinfo {author} {\bibfnamefont {T.~T.}\ \bibnamefont
  {Koutserimpas}}\ and\ \bibinfo {author} {\bibfnamefont {F.}~\bibnamefont
  {Monticone}},\ }\bibfield  {title} {\bibinfo {title} {Time-varying media,
  dispersion, and the principle of causality [{Invited}]},\ }\href@noop {}
  {\bibfield  {journal} {\bibinfo  {journal} {Optical Materials Express}\
  }\textbf {\bibinfo {volume} {14}},\ \bibinfo {pages} {1222} (\bibinfo {year}
  {2024})}\BibitemShut {NoStop}%
\bibitem [{\citenamefont {Goldstein}\ \emph {et~al.}(2002)\citenamefont
  {Goldstein}, \citenamefont {Poole}, \citenamefont {Safko},\ and\
  \citenamefont {Addison}}]{goldstein_classical_2002}%
  \BibitemOpen
  \bibfield  {author} {\bibinfo {author} {\bibfnamefont {H.}~\bibnamefont
  {Goldstein}}, \bibinfo {author} {\bibfnamefont {C.}~\bibnamefont {Poole}},
  \bibinfo {author} {\bibfnamefont {J.}~\bibnamefont {Safko}},\ and\ \bibinfo
  {author} {\bibfnamefont {S.~R.}\ \bibnamefont {Addison}},\ }\bibfield
  {title} {\bibinfo {title} {Classical {Mechanics}, 3rd ed.},\ }\href@noop {}
  {\bibfield  {journal} {\bibinfo  {journal} {American Journal of Physics}\
  }\textbf {\bibinfo {volume} {70}},\ \bibinfo {pages} {782} (\bibinfo {year}
  {2002})}\BibitemShut {NoStop}%
\end{thebibliography}%
